\documentclass[12pt,aps,prd,showpacs,superscriptaddress,nofootinbib]{revtex4-2}
\pdfoutput=1
\usepackage{braket,bm}
\usepackage[pass]{geometry} 
\usepackage{color,graphicx}
\usepackage{cancel}
\usepackage{caption}
\usepackage{subcaption}
\usepackage{float}  
\usepackage[colorlinks,citecolor=blue,linkcolor=black]{hyperref} 
\usepackage{bm}        
\usepackage{amssymb}   
\usepackage[normalem]{ulem}
\usepackage{mathrsfs,mathtools, wasysym}
\usepackage[section]{placeins} 
\DeclareMathAlphabet{\mathpzc}{OT1}{pzc}{m}{it}

\usepackage{ushort}
\usepackage{physics} 
\usepackage{accents}
\usepackage{tikz}
\usepackage{tikz-feynman} 
\tikzfeynmanset{compat=1.1.0}

\newcommand{\ul}[1]{\underaccent{\bar}{#1}}
\newcommand{\uc}[1]{\underaccent{\circ}{#1}}
\newcommand{\ub}[1]{\underaccent{\bullet}{#1}}
\newcommand{\D}{\mathrm{d}}
\newcommand\numberthis{\addtocounter{equation}{1}\tag{\theequation}}

\begin{document}

\title{Towards a Manifestly Causal Approach to Particle Scattering}
\author{Robert Dickinson}
\author{Jeff Forshaw}
\author{Ross Jenkinson\footnote{Author to whom any correspondence should be addressed.}}
\affiliation{Department of Physics and Astronomy,
  University of Manchester,
  Manchester M13 9PL,
  United Kingdom}
\author{Peter Millington}
\affiliation{Department of Physics and Astronomy,
  University of Manchester,
  Manchester M13 9PL,
  United Kingdom}

\date{\today}

\begin{abstract}
We introduce a new, probability-level approach to calculations in scalar field particle scattering. The approach involves the implicit summation over final states, which makes causality manifest since retarded propagators emerge naturally. Novel diagrams represent algebraic terms at the probability level, akin to Feynman diagrams at the amplitude level. We conjecture a list of rules that generate all probability-level diagrams for particle scattering processes in which one is fully inclusive over final states that contain no initial-state particles. These rules are confirmed using some fixed-order examples. The inclusivity and causal structure of this formalism may offer insights into the cancellation of infrared divergences if applied to gauge theory calculations.

\end{abstract}


\maketitle


\section{Introduction}

In relativistic quantum field theories (QFTs), causality is encoded in the vanishing of the commutator (or anticommutator) of field operators. However, scattering calculations are usually performed in a manner in which causality is not manifest, due to the ubiquity of the Feynman propagator as opposed to the retarded propagator (which is manifestly causal). Furthermore, Bogoliubov's \textit{Condition of Causality}~\cite{Bogolyubov:1959bfo} shows that causality is manifest and meaningful only at the probability level. The \(S\)-matrix does not directly demonstrate causality because it represents only the transition amplitude, not the full probabilistic outcome of a process. This distinction underscores the importance of observable probabilities over amplitudes when discussing physical principles like causality.

With this in mind, a manifestly causal, probability-level formalism for calculations in QFT was developed~\cite{Dickinson:2013lsa, Dickinson:2016oiy, Dickinson:2017gtm, Dickinson:2017uit} by applying a generalisation of the Baker-Campbell-Hausdorff lemma to the transition probability, naturally expressed in terms of commutators of fields. This method has been used to study the Fermi two-atom problem~\cite{Dickinson:2017gtm} and the Unruh effect~\cite{Dickinson_2025}. It also allows to construct semi-inclusive observables that implicitly sum over all relevant final states~\cite{Dickinson:2017gtm}, having parallels with the approach of weighted cross-sections due to Ore and Sterman~\cite{ORE198093}. In this paper, we apply the formalism to particle scattering processes for inclusive final states where we demand only that there are no initial-state particles in the final state. Since all of the expressions are at the probability level, they are algebraically more complicated than in the usual amplitude-level approach. In this initial study, we start with simpler scalar field theories, which offer an instructive analogy to gauge theories such as Quantum Electrodynamics (QED) and Quantum Chromodynamics (QCD). The result is a new probability-level, diagrammatic method for calculating scattering probabilities in which the retarded propagator plays a key role, making causality explicit.

The appearance of retarded propagators may offer a novel window on infrared divergences, which arise in part due to low-energy (soft) massless particles having arbitrarily large wavelengths that have effects over infinite separations. Retarded propagators are zero for spacelike separations and carry a distinct analytic structure compared to the Feynman propagator, and this approach may help quell these infinite-distance contributions. The Bloch-Nordsieck~\cite{Bloch:1937pw} and the Kinoshita-Lee-Nauenberg (KLN)~\cite{10.1063/1.1724268, PhysRev.133.B1549} theorems ensure that, while individual amplitudes may diverge, physical observables (like cross sections and decay rates) remain finite when real and virtual soft emissions are properly summed. Our formalism allows one to sum implicitly over all relevant final states from the beginning of the calculation and thereby explore the infrared behaviour from a new angle. 

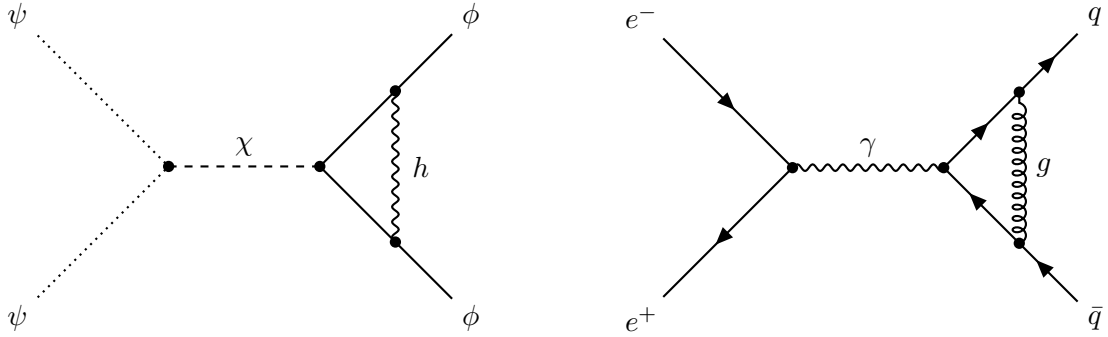
\begin{figure}[hbt]
    \centering
    \begin{tikzpicture}
        \begin{feynman}
            \vertex (i1) {$\psi$};
            \vertex[below=4cm of i1] (i2) {$\psi$};
            \vertex at ($(i1)!0.5!(i2) + (2cm, 0)$) (a);
            \vertex[right=2cm of a] (b);
            \vertex[right=6cm of i1] (f1) {$\phi$};
            \vertex[right=6cm of i2] (f2) {$\phi$};

            \vertex at ($(b) + (1cm, 1cm)$) (g1);
            \vertex at ($(b) + (1cm, -1cm)$) (g2);

            \diagram*{
                (i1) -- [ghost, thick] (a),
                (i2) -- [ghost, thick] (a),
                (a) -- [scalar, edge label=\(\chi\), thick] (b),
                (b) -- [plain, thick] (g1),
                (b) -- [plain, thick] (g2),
                (g2) -- [photon, thick, edge label'=$\:h$] (g1),
                (g1) -- [plain, thick] (f1),
                (g2) -- [plain, thick] (f2),
            };
        \end{feynman}
        \node[fill=black, circle, inner sep=1.5pt] at (a) {};
        \node[fill=black, circle, inner sep=1.5pt] at (b) {};
        \node[fill=black, circle, inner sep=1.5pt] at (g1) {};
        \node[fill=black, circle, inner sep=1.5pt] at (g2) {};
    \end{tikzpicture}
    \qquad \qquad
    \begin{tikzpicture}
        \begin{feynman}
            \vertex (i1) {$e^-$};
            \vertex[below=4cm of i1] (i2) {$e^+$};
            \vertex at ($(i1)!0.5!(i2) + (2cm, 0)$) (a);
            \vertex[right=2cm of a] (b);
            \vertex[right=6cm of i1] (f1) {$q$};
            \vertex[right=6cm of i2] (f2) {$\bar{q}$};

            \vertex at ($(b) + (1cm, 1cm)$) (g1);
            \vertex at ($(b) + (1cm, -1cm)$) (g2);

            \diagram*{
                (i1) -- [fermion, thick] (a),
                (i2) -- [anti fermion, thick] (a),
                (a) -- [photon, edge label=\(\gamma\), thick] (b),
                (b) -- [fermion, thick] (g1),
                (b) -- [anti fermion, thick] (g2),
                (g2) -- [gluon, thick, edge label'=$\:g$] (g1),
                (g1) -- [fermion, thick] (f1),
                (g2) -- [anti fermion, thick] (f2),
            };
        \end{feynman}
        \node[fill=black, circle, inner sep=1.5pt] at (a) {};
        \node[fill=black, circle, inner sep=1.5pt] at (b) {};
        \node[fill=black, circle, inner sep=1.5pt] at (g1) {};
        \node[fill=black, circle, inner sep=1.5pt] at (g2) {};
    \end{tikzpicture}
    \caption{Traditional, amplitude-level diagrams for the processes of interest.
    \textbf{Right:} The gauge theory process $e^- e^+ \rightarrow \gamma \rightarrow q \overline{q}$, with gluon ($g$) corrections.
    \textbf{Left:} An analogous toy-model in which $\psi, \chi, \phi$ and $h$ are all real scalar fields.}
    \label{fig:analogy}
\end{figure}

An example of a typical gauge theory process of interest is electron-positron ($e^+ e^-$) pair annihilation to a quark-antiquark ($q\overline{q}$) pair, mediated by a photon ($\gamma$), with gluon ($g$) corrections to the final state. The Feynman diagram for this process is pictured on the right of Fig.~\ref{fig:analogy}. A scalar analogue of this process is $\psi\psi \rightarrow \chi \rightarrow \phi\phi$ with $h$ corrections to the final state, where $\psi, \chi, \phi,$ and $h$ are all real scalar fields, as pictured on the left of Fig.~\ref{fig:analogy}. In this paper, we first consider the decay process $\chi \rightarrow X$ (Sections~\ref{sec:decay_zeroth} and~\ref{sec:firstorderdecay}), where $X$ represents a general final state, before turning to the annihilation process $\psi\psi \rightarrow X$ (Section~\ref{sec:annihilation}). For both processes, we are fully inclusive over final states, except for demanding that they contain no initial-state particles. This first study allows us to demonstrate how our new formalism plays out in a simple example, without the technical complications of spinor and gauge structure.

In Section~\ref{sec:setup}, we introduce the probability-level formalism used throughout the paper. Section~\ref{sec:rules} defines a new diagrammatic method and conjectures the general rules for generating the set of diagrams relevant to any scalar-field scattering process for which the final state contains anything except initial-state particles. For those interested in the calculational details, Sections~\ref{sec:decay_zeroth}-\ref{sec:annihilation} contain algebraic derivations of the complete sets of relevant diagrams without using the rules. We observe that distinct final states are intrinsically summed over and are not separable in our results, and the retarded self-energy emerges naturally. The calculations look complicated in the intermediate steps, but they reduce considerably. This suggests the existence of a more fundamental set of rules. Section~\ref{sec:summary} concludes.

Throughout this paper, we adopt natural units $c = \hbar = 1$ and the `mostly-minus' metric signature $(+ \,\mathord{-} \,\mathord{-} \,\mathord{-})$.

\section{Probability-Level Formulation}\label{sec:setup}

Suppose that a system is initially ($t = t_i$) described by a density operator $\rho_i = \ket{i}\bra{i}$ and that a measurement outcome is described by an effect operator $E$. In general, $E$ is an element of a Positive Operator-Valued Measure, but in the case of an exclusive final state, $\ket{f}$, it becomes a projection operator,
\begin{align} \label{eq:effect}
    E = \ket{f} \bra{f} .
\end{align} 
The probability of the measurement outcome, $\mathbb{P}$, is then given by 
\begin{equation}
\mathbb{P} \ =\  \tr (E  \rho_f)\,, \label{eq:prob}
\end{equation}
where
\begin{equation}
\rho_f \ \equiv\  U\,\rho_i \,U^\dag
\end{equation}
is the density operator at time $t_f$ and
\begin{equation}
    U = \mathrm{T} \left\{ \exp \left(\frac{1}{i} \int^{t_f}_{t_i} \D t \: \mathcal{H}_{\text{int}} (t) \right) \right\} 
\end{equation}
is the unitary evolution operator ($\mathrm{T}$ indicates time ordering) for a given interaction Hamiltonian, $\mathcal{H}_{\text{int}}(t)$.
From Eq.~\eqref{eq:prob}, the probability of the measurement outcome can be written as
\begin{equation}
    \mathbb{P} \ =\ \Braket{i| U^\dagger \: E \: U | i} .
    \label{prob}
\end{equation}
Using Eq.~\eqref{eq:effect},
\begin{equation}
    \mathbb{P} \ =\ \bra{i} U^\dagger \ket{f} \bra{f} U  \ket{i} \ =\ \big\lvert \bra{f} U \ket{i} \big\rvert^2 \,.
    \label{originalprob}
\end{equation}
This is the usual expression for the probability of an initial state, $\ket{i}$, evolving into a final state, $\ket{f}$.

Rather than calculating the amplitude and squaring, we can instead work directly with the expression in Eq.~\eqref{prob}. We do so with a generalization of the Baker-Campbell-Hausdorff lemma as in \cite{Franson_2002, Dickinson:2013lsa, Cliche_2010}. One commutes the effect operator, $E$, through the evolution operator, $U$, resulting in
\begin{alignat}{3}
    & &&\mathbb{P} &&= \sum_{j=0}^\infty \int^{t_f}_{t_i} \D t_1 \D t_2 \dots \D t_j \Theta_{1 2 \dots j} \Braket{i| \mathcal{F}_j | i} , \label{eq:newprob}\\
    \text{where }& \hspace{5mm} &&\Theta_{jk\ell\dots} &&\equiv
        \begin{cases}
            1,      & \text{if } t_j > t_k > t_\ell \dots\\
            0,      & \text{otherwise}
        \end{cases} , \label{eq:theta}\\
    \text{ }& \hspace{5mm} &&\mathcal{F}_0 &&= E , \label{F0}\\
    \text{and }& \hspace{5mm} &&\mathcal{F}_j &&= \frac{1}{i} \left[ \mathcal{F}_{j-1} , \mathcal{H}_\text{int} (t_j) \right] . \label{eq:Fj}
\end{alignat}
The recursive commutation in Eq.~\eqref{eq:Fj} leads to the direct emergence of commutators of scalar fields, which vanish for space-like separations, providing an explicit manifestation of causality.

In general, $E$ may be written as the sum of the effect operators of each final state, which themselves are products of Hermitian operators in different Hilbert spaces, i.e.,
\begin{equation}
    E = \sum_\kappa \prod_i E^{\mathscr{H}_i}_{(\kappa)} , 
\end{equation}
where the superscripts indicate the Hilbert spaces of each operator, and $\{\kappa\}$ is the set of final states. If all final states of a particular Hilbert space are summed over, then the effect operator in that Hilbert space is the identity operator, $E^{\mathscr{H}_i} = \mathbb{I}^{\mathscr{H}_i}$. If this is done for all Hilbert spaces, then our calculation is completely inclusive and calculates the probability of `anything at all' happening. It is trivial to see from Eq.~\eqref{prob} that this probability is 1, as expected. 

If the effect operator is the identity operator, Eq.~\eqref{eq:Fj} will result in nested commutators involving only functions of fields. To evaluate these, we use Eq.~(23) from Ref.~\cite{10.1063/1.1924703},
\begin{equation} \label{eq:transtrum}
    [f(\phi_1, \dots, \phi_{n-1}) , g(\phi_n)] = - \underbrace{ \sum_{k_1} \dots \sum_{k_{n-1}} } \left( \prod_{j=1}^{n-1} \frac{(- \left[\phi_j , \phi_n \right])^{k_j}}{k_j !} \right) \left(  \partial_{\phi_1}^{k_1} \dots \partial_{\phi_{n-1}}^{k_{n-1}} f \partial_{\phi_n}^{k} g \right) \,,
\end{equation}
where
\begin{equation}
    k = \sum_{j=1}^{n-1} k_{j} \,,
\end{equation}
$\phi_j \equiv \phi(x_j)$ is a field operator, and the indices within the underbrace ($^{\underbrace{\quad}}$) are not all simultaneously zero. 


\section{Diagrams and Rules}\label{sec:rules}

For the scattering processes we consider, we have developed a set of rules akin to Feynman rules, but applicable at the probability level. We take $t_i \to -\infty$ and $t_f\to \infty$ and use the usual asymptotic Fock states. Let us first explain the notation used in the diagrams. For a diagram which contributes to $\expval{\mathcal{F}_n}$, where $\langle \cdots \rangle \equiv \bra{\,i\,} \cdots \ket{\,i\,}$:
\begin{itemize}
    \item Times run from $t_n$ (earliest) on the left to $t_1$ (latest) on the right and are denoted by dots.
    \item Lines between two dots represent propagators.
    \item Lines only connected to one dot represent plane waves from the initial state at $t_i\rightarrow -\infty$ (these are drawn vertically).
    \item Feynman propagators are denoted by black lines in all Hilbert spaces.
    \item Retarded propagators are denoted by red lines in all Hilbert spaces, with an arrow pointing from the earlier time to the later time.
    \item Plane waves highlighted in yellow carry momentum $p$ out of the diagram (i.e., carry momentum $-p$ into the vertex), and unhighlighted plane waves carry momentum $p$ into the diagram.
\end{itemize}
For the scalar field theories considered, we indicate $\psi$-space contributions with dotted lines, $\chi$-space contributions with dashed lines, $\phi$-space contributions with solid lines, and $h$-space contributions with wiggly lines, as pictured in Fig.~\ref{fig:analogy}.

We can now describe the rules which can be used to construct a diagram that contributes to $\expval{\mathcal{F}_n}$. The following rules apply for a calculation which demands that there are no initial-state particles in the final state and which is fully inclusive over all other Hilbert spaces (i.e., $E^{\mathscr{H}_i}=\mathbb{I}$ in all Hilbert spaces, except for the Hilbert space of the initial-state fields, in which $E^{\mathscr{H}_i} = \ket{0} \bra{0}$). The rules are conjectured based on explicit calculations in Sections~\ref{sec:decay_zeroth}--\ref{sec:annihilation}. The rules are:
\begin{enumerate}
    \item Draw $n$ time points, which will later become vertices.
    \item Connect one initial state (regardless of the number of field quanta) to $t_1$ (the latest time)\footnote{This assumes that the Hilbert space associated with the initial state \textit{only} has particles in the initial state, i.e., the calculation is to $\order{g_i^2}$ in the coupling constant associated with the initial state Hilbert space, $g_i$.}.
    \item Connect the other initial state to any other time. 
    \item Ensure each vertex which is not connected to an initial state is connected to a later vertex by a retarded propagator of any field.
    \item Connect the remaining vertices with retarded or Feynman propagators, using the interaction vertices from the interaction Hamiltonian.
    \item If the two initial states are connected by a chain of retarded propagators (where each point on the chain connects to a later time), then momentum $p$ must flow into the $t_1$ initial state and out of the other initial state. Otherwise, choose either momentum flow.
\end{enumerate}
We highlight that, as a consequence of Rule 4, every vertex is connected to at least one of the initial states by a chain of retarded propagators. This causal structure constrains the spacetime coordinates of the interaction vertices. It is in this sense that we speak of `manifest causality'. Ref.~\cite{Dickinson:2016oiy} shows how causality explicitly plays out for initial and final states defined at fixed spacetime points. Of course, this is not possible for the scattering of infinite plane waves. 

\begin{figure}
    \centering
    \includegraphics[width=\linewidth]{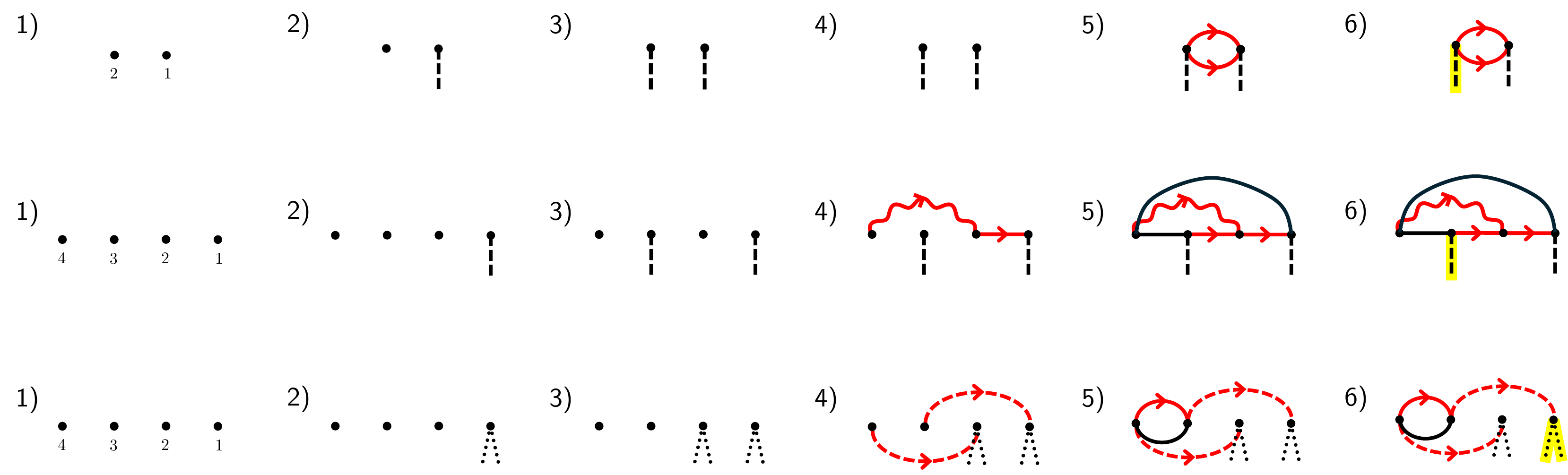}
    \caption{Examples of generating a diagram using the rules. Each step corresponds to each rule number. Feynman propagators are denoted by black lines and retarded propagators are denoted by red lines with an arrow pointing from the earlier time to the later time. \textbf{Top:} A diagram for the tree-level contribution to $\chi \rightarrow \phi\phi$. Step 4 is redundant since there are no non-initial-state vertices. The pre-factor for this diagram is $2g_\chi^2/ 2 \omega_p$. This specific diagram is generated in Section~\ref{sec:decay_zeroth}, and can be seen in Fig.~\ref{fig:treeleveldecay} (the final diagram). \textbf{Middle:} A diagram for a first-order $h$ correction to $\chi \rightarrow \phi\phi$. The pre-factor for this diagram is $16g_\chi^2 g_h^2 / 2 \omega_p$. This specific diagram is generated in Section~\ref{sec:firstorderdecay}, and can be seen in Fig.~\ref{fig:vertex} (4th line from the top, 2nd diagram from the right). \textbf{Bottom:} A diagram for the annihilation process to $\psi \psi \rightarrow X$. The pre-factor for this diagram is $16g_\chi^2 g_h^2 / 4 \omega_{p_1} \omega_{p_2}$. This specific diagram is generated in Section~\ref{sec:annihilation}, and can be seen in Fig.~\ref{fig:2to2sorted} (7th line from the top, 1st diagram from the left).}
    \label{fig:examples}
\end{figure}

The pre-factor for a diagram which contributes to $\mathbb{P}$ through $\expval{\mathcal{F}_n}$ can be calculated as follows: 
\begin{itemize}
    \item $\times 1/2\omega_i$ for each initial particle of energy $\omega_i$ in the initial state $\ket{i}$
    \item $\times Sg_i$ for each vertex, where $g_i$ is the associated coupling constant and $S$ is the symmetry factor if particles are indistinguishable
    \item $\times (-1)(-i)^n$ (these factors come from the incoming plane waves and the total number of commutators in the effect operators)
    \item $\times (-1)$ for each retarded propagator
    \item $\times 1/2$ for each loop of retarded propagators between two (and only two) vertices
    \item $\times 1/2$ for each loop of Feynman propagators between two (and only two) vertices
    \item $\times 1/2$ for each instance of a Feynman propagator connected to the same vertex at each end (Feynman propagator tadpole) 
    \item Integrate over time-ordered spacetime points, $\int \D^4x_1 \ldots \D^4x_n \Theta_{12\ldots n}$, where $\Theta_{12\ldots n}$ is defined by Eq.~\eqref{eq:theta}
\end{itemize}
The total probability, $\mathbb{P}$, is calculated by the summation of all possible diagrams. The total number of unique diagrams in Sections~\ref{sec:decay_zeroth}-\ref{sec:annihilation} matches the number determined using these rules, serving as a cross-check for their validity.
To illustrate how to construct a diagram, three examples are given in Fig.~\ref{fig:examples}.


\section{Inclusive Decay: Lowest Order} \label{sec:decay_zeroth}

\begin{figure}
    \centering
    \begin{tikzpicture}
        \begin{feynman}
            \vertex (a) {$\chi$};
            \vertex[right = 2.5cm of a] (b);
            \vertex at ($(b) + (2cm,2cm)$) (f1) {$\phi$};
            \vertex at ($(b) + (2cm,-2cm)$) (f2) {$\phi$};

            \diagram*{
                (a) -- [scalar, thick] (b),
                (b) -- [plain, thick] (f1),
                (b) -- [plain, thick] (f2),
            };
        \end{feynman}
        \node[fill=black, circle, inner sep=1.5pt] at (b) {};
    \end{tikzpicture}
    \caption{The traditional tree-level Feynman diagram for the process \(\chi\rightarrow\phi\phi\).}
    \label{fig:feynman decay tree}
\end{figure}
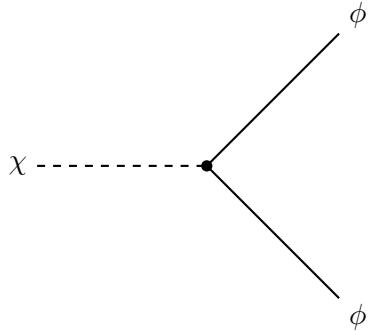

To calculate the decay probability of a particle of a scalar field, $\chi$, consider the interaction Hamiltonian,
\begin{equation} \label{generalH}
    \mathcal{H}_\text{int} (t_j) = \int \D^3{\mathbf{x}_j} \, \left( g_\chi \phi^2_{j} \chi_{j} + g_h \phi^2_{j} h_{j} \right)\,,
\end{equation}
where $\phi_j \equiv \phi(x_j)$ and $h_j \equiv h(x_j)$ are scalar fields, and $g_\chi$ and $g_h$ are coupling constants. Since $\chi$ only couples to $\phi^2$, the decay $\chi \rightarrow \phi\phi$ is the only lowest-order process which can occur. The scalar field, $h$, will be used in Section~\ref{sec:firstorderdecay} to investigate higher-order corrections. The traditional, amplitude-level Feynman diagram for the lowest-order process is shown in Fig.~\ref{fig:feynman decay tree}.

The initial state is encoded in an initial density operator
\begin{equation}
    \rho_0 = \ket{0^h\,p^\chi\,0^\phi} \bra{{0^h\,p^\chi\,0^\phi}} . \label{hchiphi:initial}
\end{equation}
This means that the system initially has one $\chi$ particle, of momentum $p$, and no other field excitations.
For the final state, we consider any number of excitations in the $\phi$ and $h$ fields and no excitations in the $\chi$ field. This is encoded in an effect operator
\begin{equation}
\begin{split}
    E = \sum_{n,\alpha} \ket{n^h\,0^\chi\,\alpha^\phi} \bra{n^h\,0^\chi\,\alpha^\phi} = \mathbb{I}^h \, \ket{0^\chi} \bra{0^\chi} \, \mathbb{I}^\phi \; , \\
    \text{i.e., } \quad E^h = \mathbb{I}^h \; , \quad E^\chi = \ket{0^\chi} \bra{0^\chi} \; , \quad E^{\phi} = \mathbb{I}^\phi \; ,
    \label{Escattering}
\end{split}
\end{equation}
where we have used the completeness of states to sum over all states in the $\phi$ and $h$ Hilbert spaces, resulting in identity operators. It is in this way that inclusive observables take a straightforward form at the probability level.

This set-up allows us to use the rules in Section~\ref{sec:rules}, but we first calculate the transition probability directly. The result can then be used to verify the rules.

If we wanted to completely reproduce the expressions of the traditional scattering calculation, we would set $E^h = \ket{0^h} \bra{0^h}$ (at $\order{g_h^0}$), $ E^\chi = \ket{0^\chi} \bra{0^\chi}$, and $E^{\phi} = \ket{q_1^\phi , q_2^\phi} \bra{q_1^\phi , q_2 ^\phi}$. The probability would then factorise into amplitude and conjugate amplitude.

Returning to the effect operator in Eq.~\eqref{Escattering}, we now define the useful notation:
\begin{alignat}{3}
    \text{Hilbert space $\phi$: }& \hspace{7mm} &&\mathcal{E}^{\dots h}_{\dots k} \coloneqq \, \frac{1}{i} \comm{\mathcal{E}^{\dots}_{\dots}}{g_h \phi^2 _k}, \hspace{7mm} &&\mathcal{E}^{\dots h}_{\dots \ul{k}} \coloneqq \,\acomm{\mathcal{E}^{\dots}_{\dots}}{g_h \phi^2_k}, \nonumber \\
    & \hspace{7mm} &&\mathcal{E}^{\dots \chi }_{\dots k} \coloneqq \,\frac{1}{i} \comm{\mathcal{E}^{\dots}_{\dots}}{g_\chi  \phi^2_k}, \hspace{7mm} &&\mathcal{E}^{\dots \chi }_{\dots \ul{k}} \coloneqq \,\acomm{\mathcal{E}^{\dots}_{\dots}}{g_\chi  \phi^2_k}, \label{curlyEdots} \\
    \text{Hilbert space $h$: }& \hspace{7mm} &&E^h_{\dots k} \coloneqq \frac{1}{i} \comm{E^h_{\dots}}{h_k}, \hspace{10mm} &&E^h_{\dots \ul{k}} \coloneqq \acomm{E^h_{\dots}}{h_k}, \nonumber \\
    \text{Hilbert space $\chi$: }& \hspace{7mm} &&E^\chi _{\dots k} \coloneqq \frac{1}{i} \comm{E^\chi _{\dots}}{\chi _k}, \hspace{10mm} &&E^\chi _{\dots \ul{k}} \coloneqq \acomm{E^\chi _{\dots}}{\chi _k}, \label{Edots}
\end{alignat}
where $\mathcal{E} \equiv E^\phi $.
The general formula for $\mathcal{F}_n$, as in \cite{Dickinson:2016oiy}, is
\begin{align}
    \mathcal{F}_n &= 2^{-n}\int \prod_{\kappa = 1}^{n} \left( \D^3 {\mathbf{x}_{\kappa}}\right) \sum_{a\,=\,0}^{n}E^h_{(\uc{1}\ldots\!\underaccent{\!\!\cdots}{}\; \uc{a}}\,E^\chi_{a+\!\uc{}\,1\ldots\!\!\!\underaccent{\cdots}{}\;\;\uc{n})}\,\mathcal{E}^{(\!h\ldots h\,\ \chi \ldots \chi )}_{(\ub{1}\,\ldots\underaccent{\!\!\!\!\cdots}{}\,\ub{a}\,a+\!\ub{}\,1\ldots\!\!\!\underaccent{\cdots}{}\;\;\ub{n})} ~,
\label{eq:Fn}
\end{align}
wherein we integrate over all spatial coordinates $\mathbf{x}_\kappa$.
This expression has been written in a condensed form by introducing an underdot notation, where, e.g.,
\begin{equation}
\begin{split}
    E^h_{\uc{k}} \mathcal{E}^h_{\ub{k}} \coloneqq &\; E^h_k \mathcal{E}^h_{\ul{k}} + E^h_{\ul{k}} \mathcal{E}^h_k ,
    \\
    E^h_{\uc{k} \uc{l}} \mathcal{E}^{hh}_{\ub{k} \ub{l}} \coloneqq &\;
    E^h_{k l} \mathcal{E}^{hh}_{\ul{k} \ul{l}}
    + E^h_{k \ul{l}} \mathcal{E}^{hh}_{\ul{k} l}
    + E^h_{\ul{k} l} \mathcal{E}^{hh}_{k \ul{l}}
    + E^h_{\ul{k} \ul{l}} \mathcal{E}^{hh}_{k l} ,
    \\
    E^h_{\uc{k}} E^\chi_{\uc{l}} \mathcal{E}^{h\chi}_{\ub{k} \ub{l}} \coloneqq &\; E^h_k E^\chi_l \mathcal{E}^{h\chi}_{\ul{k} \ul{l}} + E^h_{\ul{k}} E^\chi_l \mathcal{E}^{h\chi}_{k \ul{l}} + E^h_k E^\chi_{\ul{l}} \mathcal{E}^{h\chi}_{\ul{k} l} + E^h_{\ul{k}} E^\chi_{\ul{l}} \mathcal{E}^{h\chi}_{k l} ,
\end{split}
\end{equation}
which encodes a sum over all possible permutations of commutation/anticommutation pairs. An index with underdots must be underlined once, and only once, across the operators from different Hilbert spaces. We then sum over the possibilities. The parentheses indicate a summation over all permutations of those indices which result in unique terms (remembering that the time indices are time-ordered within each operator). For example, we need not consider $E^h_{\uc{3}\uc{1}}E^\chi_{\uc{2}}\mathcal{E}^{h\chi h}_{\ub{1}\ub{2}\ub{3}}$, as this is not unique to $E^h_{\uc{1}\uc{3}}E^\chi_{\uc{2}}\mathcal{E}^{h\chi h}_{\ub{1}\ub{2}\ub{3}}$.
The superscript letter indices on the $\mathcal{E}$ operators will shuffle such that they will always remain aligned with the subscript number index corresponding to the $E$ operator which also carries that number index. 
An example of expanding the parentheses is given below:
\begin{equation}
    E^h_{(\uc{1}\uc{2}}E^\chi_{\uc{3})}\mathcal{E}^{(h h \chi)}_{(\ub{1}\ub{2}\ub{3})} =
    E^h_{\uc{1}\uc{2}}E^\chi_{\uc{3}}\mathcal{E}^{hh \chi}_{\ub{1}\ub{2}\ub{3}} +
    E^h_{\uc{1}\uc{3}}E^\chi_{\uc{2}}\mathcal{E}^{h \chi h}_{\ub{1}\ub{2}\ub{3}} +
    E^h_{\uc{2}\uc{3}}E^\chi_{\uc{1}}\mathcal{E}^{\chi hh}_{\ub{1}\ub{2}\ub{3}}\,.
\end{equation}

The lowest order non-zero contribution to $\mathbb{P}$ is from $\mathcal{F}_2$. Using Eq.~\eqref{eq:Fn},
\begin{align}
    \mathcal{F}_2 &= \frac{1}{4} \int \D^3 {\mathbf{x}_{1}} \D^3 {\mathbf{x}_{2}} \left( E^h_{\uc{1} \uc{2}}\,E^\chi \,\mathcal{E}^{h h}_{\ub{1} \ub{2}}
    + E^h_{\uc{1}}\,E^\chi_{\uc{2}}\,\mathcal{E}^{h \chi}_{\ub{1} \ub{2}}
    + E^h_{\uc{2}}\,E^\chi_{\uc{1}}\,\mathcal{E}^{\chi h}_{\ub{1} \ub{2}}
    + E^h \,E^\chi_{\uc{1} \uc{2}}\,\mathcal{E}^{\chi \chi}_{\ub{1} \ub{2}} \right) ~.
\label{eq:F2}
\end{align}
Due to our choices of the initial density operator, $\rho_0$, and the effect operator, $E$, only the final term contributes once we take the expectation value of $\mathcal{F}_2$, as in Eq.~\eqref{eq:newprob}. The first subscript index of $\mathcal{E}^{\dots}_{\dots}$ must be underlined, since a commutator with $\mathcal{E} = \mathbb{I}^\phi$ would vanish. Thus, the only contribution from Eq.~\eqref{eq:F2} is
\begin{align}
    \mathcal{F}_2 &= \frac{1}{4} \int \D^3 {\mathbf{x}_{1}} \D^3 {\mathbf{x}_{2}} \,E^\chi_{1 \uc{2}}\,\mathcal{E}^{\chi \chi}_{\ul{1} \ub{2}} \nonumber\\
    &= \frac{1}{4} \int \D^3 {\mathbf{x}_{1}} \D^3 {\mathbf{x}_{2}} \,\left( E^\chi_{1 2}\,\mathcal{E}^{\chi \chi}_{\ul{1} \ul{2}} + E^\chi_{1 \ul{2}}\,\mathcal{E}^{\chi \chi}_{\ul{1} 2} \right)
    ~.
\label{eq:relevantF2}
\end{align}
Therefore,
\begin{align} \label{eq:F2expval}
    \bra{i} \mathcal{F}_2 \ket{i} &= \frac{1}{4} \int \D^3 {\mathbf{x}_{1}} \D^3 {\mathbf{x}_{2}} \,\left( \bra{p^\chi} E^\chi_{1 2} \ket{p^\chi} \, \bra{0^\phi} \mathcal{E}^{\chi \chi}_{\ul{1} \ul{2}} \ket{0^\phi} + \bra{p^\chi} E^\chi_{1 \ul{2}} \ket{p^\chi} \, \bra{0^\phi} \mathcal{E}^{\chi \chi}_{\ul{1} 2} \ket{0^\phi} \right)
    ~.
\end{align}
Evaluating the $\chi$-space expectation value first gives
\begin{align}
    \bra{p^\chi} E^\chi_{1 \uc{2}} \ket{p^\chi} &= \bra{0^\chi} a(p) \, E^\chi_{1 \uc{2}} \,a^\dagger(p) \ket{0^\chi} \nonumber\\
    &= \frac{1}{i} \biggl(\frac{1}{i}\biggr)^{(1-\eta_2)/2} \bra{0^\chi} a(p) \, \biggl[ \Bigl[ \ket{0^\chi}\bra{0^\chi} , \chi_1 \Bigr], \chi_2 \biggr]_{\eta_2} \,a^\dagger(p) \ket{0^\chi} \nonumber\\
    &= - \biggl(\frac{1}{i}\biggr)^{(3-\eta_2)/2} \bra{0^\chi} a(p) \, \chi_1  \ket{0^\chi}\bra{0^\chi} \chi_2 \, a^\dagger(p) \ket{0^\chi} \nonumber \\
    & \qquad \qquad \qquad \qquad + \eta_2 \, \bra{0^\chi} a(p) \, \chi_2 \ket{0^\chi}\bra{0^\chi} \chi_1  \, a^\dagger(p) \ket{0^\chi}\,,
\end{align}
where we have introduced the following compact notation for commutators and anticommutators:
\begin{align} \label{eq:generalcomm}
    \bigl[ A , B \bigr]_{\eta} &= AB + \eta BA = (1+\eta)\,AB - \eta \bigl[ A, B \bigr] \notag\\
    &= 
    \begin{cases}
        \bigl[ A , B \bigr]     & \text{if } \eta = -1 \,,\\
        \bigl\{ A , B \bigr\}      & \text{if } \eta = +1 \,,
    \end{cases} 
\end{align}
such that $\eta_2 = 1$ for $E^\chi_{1 \ul{2}}$ and $\eta_2 = -1$ for $E^\chi_{1 2}$. Using the mode expansion of the $\chi$ field,
\begin{align} \label{eq:chispaceexpval}
    \bra{p^\chi} E^\chi_{1 \uc{2}} \ket{p^\chi} 
    &= \biggl(\frac{1}{i}\biggr)^{(3-\eta_2)/2} \frac{1}{2 \omega_p} \bigl( - e^{i p \cdot x_1} \, e^{-i p \cdot x_2} + \eta_2 \, e^{i p \cdot x_2} \, e^{-i p \cdot x_1} \bigr)\,,
\end{align}
where $\omega_p = \sqrt{\mathbf{p}^2 + m_\chi^2}$ is the energy of the incoming particle.

\begin{figure}
    \centering
    \includegraphics[width=\linewidth]{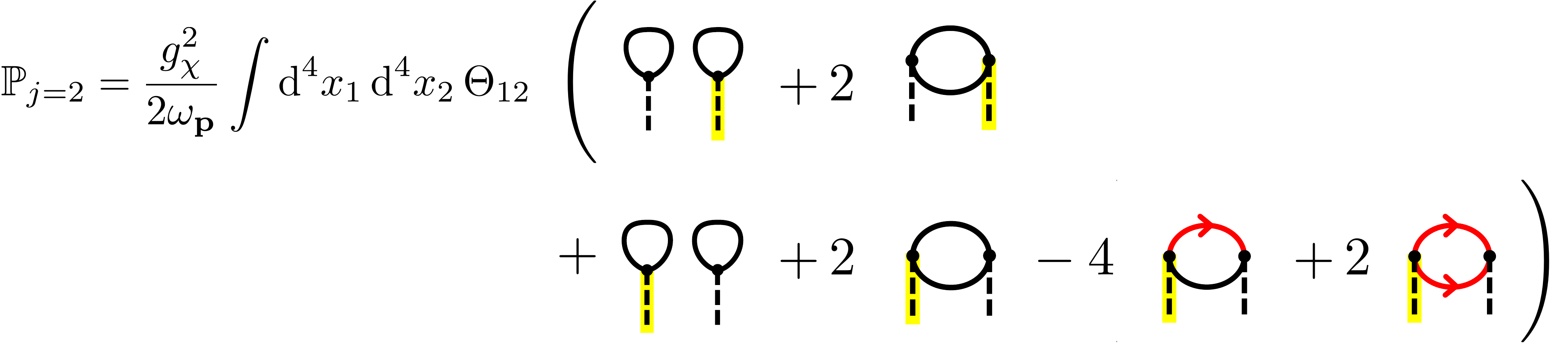}
    \caption{A diagrammatic representation of the tree-level decay probability (Eq.~\eqref{eq:probtreelevel}). }
    \label{fig:treeleveldecay}
\end{figure}

Now consider the $\phi$-space expectation value,
\begin{equation}
\begin{split}
    \bra{0^\phi} \mathcal{E}^{\chi \chi}_{\ul{1} \ub{2}} \ket{0^\phi} &= \biggl( \frac{1}{i} \biggr)^{(1-\epsilon_2) /2} g_\chi^2 \bra{0^\phi} \Bigl[ \Bigl\{ \mathbb{I}^\phi , \phi_1^2 \Bigr\}, \phi_2^2 \Bigr]_{\epsilon_2} \ket{0^\phi} \\
    &= 2 \biggl( \frac{1}{i} \biggr)^{(1-\epsilon_2) /2}\, g_\chi^2 \, \bra{0^\phi} \bigl[ \phi_1^2 , \phi_2^2 \bigr]_{\epsilon_2} \ket{0^\phi}\,,
\end{split}
\end{equation}
where $\epsilon_2 = - \eta_2$. Using Eqs.~\eqref{eq:transtrum} and~\eqref{eq:generalcomm},
\begin{align} \label{eq:phispaceexpval}
    \bra{0^\phi} \mathcal{E}^{\chi \chi}_{\ul{1} \ub{2}} \ket{0^\phi} &= \biggl( \frac{1}{i} \biggr)^{(1-\epsilon_2) /2} \biggl[ 2\,g_\chi^2\, \bra{0^\phi} (1 + \epsilon_2) \phi_1^2 \phi_2^2 \ket{0^\phi} \nonumber\\
    & \qquad \qquad \qquad - 2\,g_\chi^2\, \epsilon_2 \bra{0^\phi} \Bigl( \Delta_{12}^\phi \partial_1 \bigl[ \phi_1^2 \bigr] \partial_2 \bigl[ \phi_2^2 \bigr] - \frac{1}{2} (\Delta_{12}^\phi)^2 \partial_1^2 \bigl[ \phi_1^2 \bigr] \partial_2^2 \bigl[ \phi_2^2 \bigr] \Bigr)\ket{0^\phi} \biggr] \nonumber\\
    &= \biggl( \frac{1}{i} \biggr)^{(1-\epsilon_2) /2} \biggl[ 2 \,g_\chi^2\, (1 + \epsilon_2) \bra{0^\phi} \phi_1^2 \phi_2^2 \ket{0^\phi} \nonumber\\
    & \qquad \qquad \qquad - 8 \,g_\chi^2\, \epsilon_2 \,\Delta_{12}^\phi \bra{0^\phi} \phi_1 \phi_2 \ket{0^\phi} + 4 \,g_\chi^2\, \epsilon_2 \,(\Delta_{12}^\phi)^2 \biggr] \,,
\end{align}
where we have introduced the Pauli-Jordan function, $\Delta_{xy}^\phi$ (see Appendix~\ref{app:propagators}).
Substituting Eqs.~\eqref{eq:chispaceexpval} and~\eqref{eq:phispaceexpval} into Eq.~\eqref{eq:F2expval}, we have
\begin{align}
    \bra{i} \mathcal{F}_2 \ket{i} 
    &= \frac{g_\chi^2}{2 \omega_p} \int \D^3 {\mathbf{x}_{1}} \D^3 {\mathbf{x}_{2}} \,\bigg( 
    e^{i p \cdot x_1} \, e^{-i p \cdot x_2} \bra{0^\phi} \phi_1^2 \phi_2^2 \ket{0^\phi} \nonumber\\
    & \qquad + \, e^{i p \cdot x_2} \, e^{-i p \cdot x_1} \Bigl( \bra{0^\phi} \phi_1^2 \phi_2^2 \ket{0^\phi} -4 \,\Delta_{12}^\phi \bra{0^\phi} \phi_1 \phi_2 \ket{0^\phi} + 2\, (\Delta_{12}^\phi)^2 \Bigr) \bigg)
    ~. \label{eq:F2expvalfinal}
\end{align}
Since Eq.~\eqref{eq:newprob} includes $\Theta_{12}$, we can write
\begin{equation}
    \Theta_{12} \bra{0^\phi} \phi_1^n \phi_2^m \ket{0^\phi} = \Theta_{12} \bra{0^\phi} \mathrm{T} \{ \phi_1 ^n \phi_2^m \} \ket{0^\phi}\,, 
\end{equation}
where $n$ and $m$ are positive integers, and use Wick's Theorem to obtain
\begin{align} \label{eq:probtreelevel}
    \mathbb{P}_{j=2} &= \frac{g_\chi^2}{2 \omega_p} \int \D^4 {x_{1}} \D^4 {x_{2}} \,\Theta_{12}\,\bigg( 
     e^{i p \cdot x_1} \, e^{-i p \cdot x_2} \Bigl( F_{11}^\phi F_{22}^\phi + 2 \,F_{12}^\phi F_{12}^\phi \Bigr) \nonumber\\
    & \qquad + \, e^{i p \cdot x_2} \, e^{-i p \cdot x_1} \Bigl( F_{11}^\phi F_{22}^\phi + 2 \,F_{12}^\phi F_{12}^\phi -4 \,R_{12}^\phi F_{12}^\phi + 2\, (R_{12}^\phi)^2 \Bigr) \bigg)
    ~,
\end{align}
where we have introduced the Feynman propagator, $F_{xy}^\phi$, and the retarded propagator, $R_{xy}^\phi$ (see Appendix~\ref{app:propagators}).
Eq.~\eqref{eq:probtreelevel} is expressed diagrammatically in Fig.~\ref{fig:treeleveldecay}. These are exactly the diagrams which would be generated using the rules in Section~\ref{sec:rules}.

Since
\begin{align}
    \Theta_{12} \,F_{12}^\phi &= \Theta_{12} \,\Delta_{12}^{\phi(>)} \,,\\ \Theta_{12} \,R_{12}^\phi &= \Theta_{12} \,\Delta_{12}^\phi = \Theta_{12}^\phi \left( \Delta_{12}^{\phi(>)} - \Delta_{12}^{\phi(<)} \right)\,,
\end{align}
where the Wightman functions, $\Delta_{xy}^{\phi(>)} $ and $ \Delta_{xy}^{\phi(<)}  $, are defined in Appendix~\ref{app:propagators}, we can rewrite Eq.~\eqref{eq:probtreelevel} as
\begin{align} 
    \mathbb{P}_{j=2}
    &= \frac{2g_\chi^2}{2 \omega_p} \int \D^4 {x_{1}} \D^4 {x_{2}} \,\Theta_{12}\,\bigg( 
     e^{i p \cdot x_1} \, e^{-i p \cdot x_2} \,\Bigl( \Delta_{12}^{\phi(>)} \Bigr)^2 +  e^{i p \cdot x_2} \, e^{-i p \cdot x_1} \Bigl( \Delta_{12}^{\phi(<)} \Bigr)^2 \bigg)
    ~.
\end{align}
We have ignored the $F_{11}^\phi F_{22}^\phi$ terms since these contributions are not allowed kinematically (alternatively, we could have chosen $E^\phi = \mathbb{I}^\phi - \ket{0^\phi}\bra{0^\phi}$ such that these contributions would cancel). The part of the integrand after the $\Theta$-function is symmetric under the exchange of $t_1 \leftrightarrow t_2$, so we can replace the $\Theta$-function with $(1/2!)$. 
By substituting in the momentum-integral representations of the Wightman functions, the decay probability can then be expressed as
\begin{align} \label{eq:treelevelprobfinal}
    \mathbb{P}_{j=2} 
    = \frac{2g_\chi^2}{2 \omega_p} &
    \int \frac{\dd[3]{\mathbf{q}_1}}{(2\pi)^3} \frac{1}{2 \omega_{q_1}} \int \frac{\dd[3]{\mathbf{q}_2}}{(2\pi)^3} \frac{1}{2 \omega_{q_2}} \bigl[ \delta^4 (p - q_1 - q_2) \bigr]^2 (2\pi)^8
    ~.
\end{align}
This is the usual tree-level probability for scalar particle decay, integrated over all final state momenta.


\section{Inclusive Decay: First Order}\label{sec:firstorderdecay}

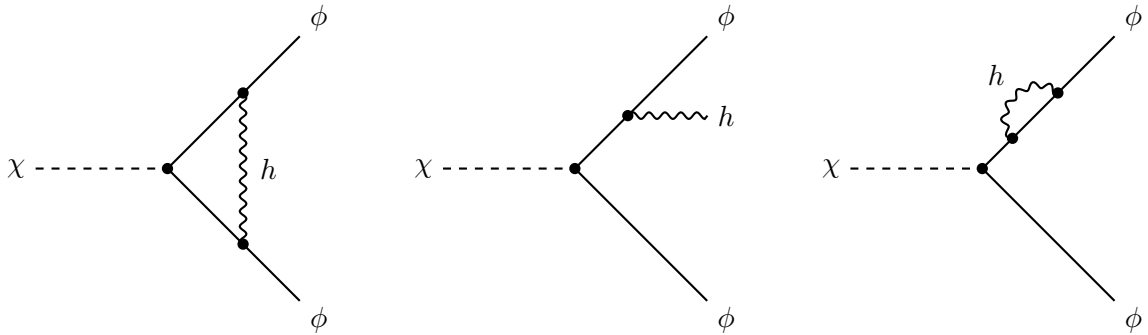
\begin{figure}
    \centering
    \begin{tikzpicture}
        \begin{feynman}
            \vertex (a) {$\chi$};
            \vertex[right = 2cm of a] (b);
            \vertex at ($(b) + (2cm,2cm)$) (f1) {$\phi$};
            \vertex at ($(b) + (2cm,-2cm)$) (f2) {$\phi$};
            \vertex at ($(b) + (1cm,1cm)$) (i1);
            \vertex at ($(b) + (1cm,-1cm)$) (i2);

            \diagram*{
                (a) -- [scalar, thick] (b),
                (b) -- [plain, thick] (f1),
                (b) -- [plain, thick] (f2),
                (i1) -- [photon, thick, edge label = $\:h$] (i2),
            };
        \end{feynman}
        \node[fill=black, circle, inner sep=1.5pt] at (b) {};
        \node[fill=black, circle, inner sep=1.5pt] at (i1) {};
        \node[fill=black, circle, inner sep=1.5pt] at (i2) {};
    \end{tikzpicture}
    \qquad
    \begin{tikzpicture}
        \begin{feynman}
            \vertex (a) {$\chi$};
            \vertex[right = 2cm of a] (b);
            \vertex at ($(b) + (2cm,2cm)$) (f1) {$\phi$};
            \vertex at ($(b) + (2cm,-2cm)$) (f2) {$\phi$};
            \vertex at ($(b) + (0.7cm,0.7cm)$) (i1);
            \vertex at ($(b) + (2cm,0.7cm)$) (i2) {$h$};

            \diagram*{
                (a) -- [scalar, thick] (b),
                (b) -- [plain, thick] (f1),
                (b) -- [plain, thick] (f2),
                (i1) -- [photon, thick] (i2),
            };
        \end{feynman}
        \node[fill=black, circle, inner sep=1.5pt] at (b) {};
        \node[fill=black, circle, inner sep=1.5pt] at (i1) {};
    \end{tikzpicture}
    \qquad
    \begin{tikzpicture}
        \begin{feynman}
            \vertex (a) {$\chi$};
            \vertex[right = 2cm of a] (b);
            \vertex at ($(b) + (2cm,2cm)$) (f1) {$\phi$};
            \vertex at ($(b) + (2cm,-2cm)$) (f2) {$\phi$};
            \vertex at ($(b) + (0.4cm,0.4cm)$) (i1);
            \vertex at ($(b) + (1cm,1cm)$) (i2);

            \diagram*{
                (a) -- [scalar, thick] (b),
                (b) -- [plain, thick] (f1),
                (b) -- [plain, thick] (f2),
                (i1) -- [photon, thick, edge label=$h$, half left] (i2),
            };
        \end{feynman}
        \node[fill=black, circle, inner sep=1.5pt] at (b) {};
        \node[fill=black, circle, inner sep=1.5pt] at (i1) {};
        \node[fill=black, circle, inner sep=1.5pt] at (i2) {};
    \end{tikzpicture}
    \caption{The Feynman diagrams for the process \(\chi\rightarrow\phi\phi\) with first-order \(h\) corrections.}
    \label{fig:feynman decay first}
\end{figure}

In this section, we will generate diagrams which represent the first-order $h$ corrections to the decay probability in Section~\ref{sec:decay_zeroth}. As such, the Hamiltonian, initial state, and effect operator remain the same as Section~\ref{sec:decay_zeroth}. Some examples of the traditional, amplitude-level Feynman diagrams for this process are shown in Fig.~\ref{fig:feynman decay first}. 

To calculate these corrections, we require
\begin{align}
    \mathcal{F}_4 &= \frac{1}{16} \int \D^3 {\mathbf{x}_{1}} \,\D^3 {\mathbf{x}_{2}} \,\D^3 {\mathbf{x}_{3}} \,\D^3 {\mathbf{x}_{4}} \sum_{a\,=\,0}^{4}E^h_{(\uc{1}\ldots\!\underaccent{\!\!\cdots}{}\; \uc{a}}\,E^\chi_{a+\!\uc{}\,1\ldots\!\!\!\underaccent{\cdots}{}\;\;\uc{4})}\,\mathcal{E}^{(\!h\ldots h\,\ \chi \ldots \chi )}_{(\ub{1}\,\ldots\underaccent{\!\!\!\!\cdots}{}\,\ub{a}\,a+\!\ub{}\,1\ldots\!\!\!\underaccent{\cdots}{}\;\;\ub{4})} ~.
\label{eq:F4}
\end{align}
The relevant term is of order $\order{g_h^2 g_\chi^2}$, which is the $a=2$ term,
\begin{align}
    \mathcal{F}^{(a=2)}_4 &= \frac{1}{16} \int \D^3 {\mathbf{x}_{1}} \,\D^3 {\mathbf{x}_{2}} \,\D^3 {\mathbf{x}_{3}} \,\D^3 {\mathbf{x}_{4}} E^h_{(\uc{1}\uc{2}}\,E^\chi_{\uc{3}\uc{4})}\,\mathcal{E}^{(h h \chi \chi )}_{(\ub{1}\ub{2} \ub{3}\ub{4})} ~.
\label{eq:F4a=2}
\end{align}
For clarity and simplicity, we will write a term with a general permutation of indices as
\begin{equation}
E^h_{\uc{i}\uc{j}}\,E^\chi_{\uc{k}\uc{l}}\,\mathcal{E}^{(h h \chi \chi )}_{\ub{1}\ub{2} \ub{3}\ub{4}} ~.
\end{equation}
Since $E^h = \mathbb{I}^h$ and $E^\phi = \mathbb{I}^\phi$, the first index of the $\phi$-space and $h$-space operators must be underlined (a commutator with the identity operator is zero):
\begin{equation}
E^h_{\ul{i}\uc{j}}\,E^\chi_{\uc{k}\uc{l}}\,\mathcal{E}^{(h h \chi \chi )}_{\ul{1}\ub{2} \ub{3}\ub{4}} ~.
\end{equation}
Now that the index `1' is underlined on the $\phi$-space operator, it cannot be underlined on either of the other operators. This means that $i \neq 1$. However, the index 1 must be the first index of an operator, so $k=1$:
\begin{equation} \label{eq:F4reducedE}
E^h_{\ul{i}\uc{j}}\,E^\chi_{1\uc{l}}\,\mathcal{E}^{(h h \chi \chi )}_{\ul{1}\ub{2} \ub{3}\ub{4}} ~.
\end{equation}
In total, there will always be four non-underlined indices. From Eqs. \eqref{curlyEdots} and \eqref{Edots}, we know that each non-underlined index results in a factor of $1/i$. Since $(1/i)^4 = 1$, we will simply ignore these factors of $i$ in this section. The expanded expectation value of Eq.~\eqref{eq:F4reducedE} is
\begin{align} \label{eq:F4expvalexpanded}
    \langle E^h_{\ul{i}\uc{j}}\,E^\chi_{1\uc{l}}\,\mathcal{E}^{(h h \chi \chi )}_{\ul{1}\ub{2} \ub{3}\ub{4}} \rangle &= 
    \langle E^h_{\ul{2}3}\,E^\chi_{14}\,\mathcal{E}^{\chi h h \chi}_{\ul{1}2 \ul{3}\ul{4}} \rangle + 
    \langle E^h_{\ul{2}3}\,E^\chi_{1\ul{4}}\,\mathcal{E}^{\chi h h \chi}_{\ul{1}2 \ul{3}4} \rangle +
    \langle E^h_{\ul{2}\ul{3}}\,E^\chi_{14}\,\mathcal{E}^{\chi h h \chi}_{\ul{1}2 3\ul{4}} \rangle +
    \langle E^h_{\ul{2}\ul{3}}\,E^\chi_{1\ul{4}}\,\mathcal{E}^{\chi h h \chi}_{\ul{1}2 3 4} \rangle
    \nonumber\\
    & + \langle E^h_{\ul{2}4}\,E^\chi_{13}\,\mathcal{E}^{\chi h \chi h}_{\ul{1}2 \ul{3}\ul{4}} \rangle + 
    \langle E^h_{\ul{2}4}\,E^\chi_{1\ul{3}}\,\mathcal{E}^{\chi h \chi h}_{\ul{1}2 3\ul{4}} \rangle +
    \langle E^h_{\ul{2}\ul{4}}\,E^\chi_{13}\,\mathcal{E}^{\chi h \chi h}_{\ul{1}2 \ul{3}4} \rangle +
    \langle E^h_{\ul{2}\ul{4}}\,E^\chi_{1\ul{3}}\,\mathcal{E}^{\chi h h \chi}_{\ul{1}2 3 4} \rangle 
    \nonumber\\
    & + \langle E^h_{\ul{3}4}\,E^\chi_{12}\,\mathcal{E}^{\chi \chi h h }_{\ul{1}\ul{2} 3\ul{4}} \rangle + 
    \langle E^h_{\ul{3}4}\,E^\chi_{1\ul{2}}\,\mathcal{E}^{\chi \chi h h }_{\ul{1}2 3\ul{4}} \rangle +
    \langle E^h_{\ul{3}\ul{4}}\,E^\chi_{12}\,\mathcal{E}^{\chi \chi h h }_{\ul{1}\ul{2} 3 4} \rangle +
    \langle E^h_{\ul{3}\ul{4}}\,E^\chi_{1\ul{2}}\,\mathcal{E}^{\chi \chi h h }_{\ul{1}2 3 4} \rangle\,.
\end{align}

We can evaluate the expectation values one Hilbert space at a time. The expectation value of the $h$-space operator is straightforward:
\begin{align} \label{eq:hspaceexpvalF4}
    \bra{0^h} E^h_{\ul{i}\uc{j}} \ket{0^h} &= \bra{0^h} \Bigl[ \Bigl\{ \mathbb{I}^h , h_i \Bigr\}, h_j \Bigr]_{\lambda_j} \ket{0^h} = 2 \, \bra{0^h} \bigl[ h_i , h_j \bigr]_{\lambda_j} \ket{0^h} \,.
\end{align}
This is either proportional to a Pauli-Jordan function ($\lambda_j = -1$) or a Hadamard function ($\lambda_j = +1$) (see Appendix~\ref{app:propagators}).
We can evaluate the expectation value of the $\chi$-space operator as we did in Section~\ref{sec:decay_zeroth}, giving
\begin{align} \label{eq:chispaceexpvalF4}
    \bra{p^\chi} E^\chi_{1 \uc{l}} \ket{p^\chi} &= \frac{1}{2 \omega_p} \bigl( - e^{i p \cdot x_1} \, e^{-i p \cdot x_l} + \eta_l \, e^{i p \cdot x_l} \, e^{-i p \cdot x_1} \bigr)\,.
\end{align}
The $\phi$-space operator in Eq.~\eqref{eq:F4reducedE} is more complicated. It can be written as
\begin{equation} \label{eq:E1234definition}
    \mathcal{E}_{\ul{1}\ub{2} \ub{3}\ub{4}} = 
    \comm{\mathcal{E}_{\ul{1}\ub{2} \ub{3}}}{\phi_4^2}_{\epsilon_4}\,,
\end{equation}
where the Hilbert spaces are no longer denoted, and we understand that there will be an overall factor of $g_h^2 g_\chi^2$. We shall therefore first consider 
\begin{equation}
    \mathcal{E}_{\ul{1}\ub{2} \ub{3}} = 
    \comm{\mathcal{E}_{\ul{1}\ub{2}}}{\phi_3^2}_{\epsilon_3} = (1+\epsilon_3) \,\mathcal{E}_{\ul{1}\ub{2}} \,\phi_3^2 - \epsilon_3 \comm{\mathcal{E}_{\ul{1}\ub{2}}}{\phi_3^2}\,,
\end{equation}
where we have temporarily ignored coupling constants, and use Eqs.~\eqref{eq:transtrum} and \eqref{eq:phispaceexpval} to get
\begin{align}
    \mathcal{E}_{\ul{1}\ub{2} \ub{3}} &=
    2\,(1+\epsilon_3)\,\Bigl(\,(1+\epsilon_2)\,\phi_1^2 \, \phi_2^2 \, \phi_3^2 - 4 \epsilon_2 \Delta_{12} \phi_1 \phi_2 \phi_3^2 + 2 \epsilon_2\, (\Delta_{12})^2 \phi_3^2 \Bigr) \nonumber\\
    & \hspace{5mm} + 4 \epsilon_3 \biggl( - 2\,\Delta_{13} \Bigl( (1+\epsilon_2) \phi_1 \, \phi_2^2 \, \phi_3 - 2\,\epsilon_2 \Delta_{12} \phi_2 \, \phi_3 \Bigr) \nonumber\\
    & \hspace{17mm} - 2\,\Delta_{23} \Bigl( (1+\epsilon_2) \phi_1^2 \, \phi_2 \, \phi_3 - 2\,\epsilon_2 \Delta_{12} \phi_1 \, \phi_3 \Bigr) \nonumber\\
    & \hspace{17mm} + 4 \Delta_{13} \Delta_{23} \Bigl( (1+\epsilon_2) \Delta_{13}^2 \phi_2^2 - \epsilon_2 \Delta_{12}
    \Bigr)\nonumber\\
    & \hspace{17mm} + (1+\epsilon_2) (\Delta_{13})^2 \,\phi_2^2 - (1+\epsilon_2) (\Delta_{23})^2 \,\phi_1^2 \biggr) \,,
\end{align}
where $\Delta_{xy} \equiv \Delta_{xy}^{\phi}$. Substituting this into Eq.~\eqref{eq:E1234definition} and using Eq.~\eqref{eq:transtrum},
{\allowdisplaybreaks
\begin{align*} \label{eq:E1234general}
    \mathcal{E}_{\ul{1}\ub{2} \ub{3}\ub{4}} &=
    g_h^2 g_\chi^2\,\Biggl\{ 2\,(1+\epsilon_4) \Biggl( (1+\epsilon_3) \biggl[ (1+\epsilon_2) \,\phi_1^2 \,\phi_2^2 \,\phi_3^2 \,\phi_4^2 - 4\,\epsilon_2 \Delta_{12} \,\phi_1 \,\phi_2 \,\phi_3^2 \,\phi_4^2 + 2 \epsilon_2 \,(\Delta_{12})^2 \,\phi_3^2 \,\phi_4^2 \biggr]  \\
    & \hspace{22mm} + 2 \epsilon_3 \biggl[ - 2\,\Delta_{13} \Bigl( (1+\epsilon_2) \,\phi_1 \,\phi_2^2 \,\phi_3 \,\phi_4^2 - 2\, \epsilon_2 \,\Delta_{12} \,\phi_2 \,\phi_3 \,\phi_4^2 \Bigr)  \\ 
    & \hspace{34mm} - 2\,\Delta_{23} \Bigl( (1+\epsilon_2) \,\phi_1^2 \,\phi_2 \,\phi_3 \,\phi_4^2 - 2 \,\epsilon_2 \,\Delta_{12} \,\phi_1 \,\phi_3 \,\phi_4^2 \Bigr)  \\
    & \hspace{34mm} + 4\,\Delta_{13}\,\Delta_{23} \Bigl( (1+\epsilon_2) \,\phi_1\,\phi_2\,\phi_4^2  - \epsilon_2 \,\Delta_{12}\,\phi_4^2\Bigr)  \\
    & \hspace{34mm} + (1+\epsilon_2) (\Delta_{13})^2 \,\phi_2^2 \,\phi_4^2 + (1+\epsilon_2) \, (\Delta_{23})^2 \phi_1^2 \,\phi_4^2 \biggr]
    \Biggr)  \\
    & \hspace{5mm} +4\, \epsilon_4 \Biggl( - 2\,\Delta_{14} 
    \biggl[ (1+\epsilon_3)\Bigl( (1+\epsilon_2)\,\phi_1\,\phi_2^2\,\phi_3^2\,\phi_4 - 2\,\epsilon_2\,\Delta_{12} \,\phi_2\,\phi_3^2\,\phi_4 \Bigr)  \\
    & \hspace{32mm} +2\,\epsilon_3 \Bigl(- (1+\epsilon_2)\,\Delta_{13}\,\phi_2^2\,\phi_3\,\phi_4  \\
    & \hspace{44mm} - 2\,\Delta_{23}\bigl((1+\epsilon_2)\,\phi_1\,\phi_2\,\phi_3\,\phi_4- \epsilon_2\,\Delta_{12}\,\phi_3\,\phi_4\bigr) \\
    & \hspace{44mm} +2\,(1+\epsilon_2) \,\Delta_{13}\,\Delta_{23}\,\phi_2\,\phi_4 + (1+\epsilon_2) \,(\Delta_{23})^2\,\phi_1 \,\phi_4 \Bigr)
    \biggr]  \\
    & \hspace{19mm} - 2\,\Delta_{24} 
    \biggl[ (1+\epsilon_3)\Bigl((1+\epsilon_2)\,\phi_1^2\,\phi_2\,\phi_3^2\,\phi_4 - 2\,\epsilon_2\,\Delta_{12} \,\phi_1\,\phi_3^2\,\phi_4 \Bigr)  \\
    & \hspace{32mm} + 2\,\epsilon_3 \Bigl( - (1+\epsilon_2)\,\Delta_{23}\,\phi_1^2\,\phi_3\,\phi_4  \\
    & \hspace{44mm} - 2\,\Delta_{13}\bigl((1+\epsilon_2)\,\phi_1\,\phi_2\,\phi_3\,\phi_4- \epsilon_2\,\Delta_{12}\,\phi_3\,\phi_4\bigr)\\
    & \hspace{44mm} +2\,(1+\epsilon_2) \,\Delta_{13}\,\Delta_{23}\,\phi_1\,\phi_4 + (1+\epsilon_2) \,(\Delta_{13})^2\,\phi_2 \,\phi_4 \Bigr)
    \biggr] \\
    & \hspace{19mm} - 2\,\Delta_{34} 
    \biggl[ (1+\epsilon_3)\Bigl( (1+\epsilon_2)\,\phi_1^2\,\phi_2^2\,\phi_3\,\phi_4 - 4\,\epsilon_2\,\Delta_{12} \,\phi_1\,\phi_2\,\phi_3\,\phi_4 + 2\, \epsilon_2\, (\Delta_{12})^2 \,\phi_3 \,\phi_4 \Bigr) \\
    & \hspace{32mm} + 2\,\epsilon_3 \Bigl( - \Delta_{13}\bigl( (1+\epsilon_2)\,\phi_1\,\phi_2^2\,\phi_4- 2\,\epsilon_2\,\Delta_{12}\,\phi_2\,\phi_4\bigr) \\
    & \hspace{44mm} - \Delta_{23}\bigl( (1+\epsilon_2)\,\phi_1^2\,\phi_2\,\phi_4- 2\,\epsilon_2\,\Delta_{12}\,\phi_1\,\phi_4\bigr) \Bigr)
    \biggr] \\
    & \hspace{19mm} +4\,\Delta_{14}\,\Delta_{24} 
    \biggl[ (1+\epsilon_3) \Bigl((1+\epsilon_2)\,\phi_1\,\phi_2\,\phi_3^2 - \epsilon_2\,\Delta_{12}\,\phi_3^2 \Bigr) \\
    & \hspace{39mm} + 2\,\epsilon_3 \,(1+\epsilon_2)\Bigl( -\Delta_{13}\,\phi_2\,\phi_3 - \Delta_{23} \,\phi_1 \,\phi_3 + \,\Delta_{13} \, \Delta_{23} \Bigr)
    \biggr] \\
    & \hspace{19mm} +4\,\Delta_{14}\,\Delta_{34} 
    \biggl[ (1+\epsilon_3) \Bigl((1+\epsilon_2)\,\phi_1\,\phi_2^2\,\phi_3 - 2\,\epsilon_2\,\Delta_{12}\,\phi_2\,\phi_3 \Bigr) \\
    & \hspace{39mm} + \epsilon_3 \Bigl( - (1+\epsilon_2)\,\Delta_{13}\,\phi_2^2 - 2\,\Delta_{23} \bigl( (1+\epsilon_2) \,\phi_1 \,\phi_2 - \epsilon_2\,\Delta_{12} \bigr) \Bigr)
    \biggr] \\
    & \hspace{19mm} +4\,\Delta_{24}\,\Delta_{34} 
    \biggl[ (1+\epsilon_3) \Bigl((1+\epsilon_2)\,\phi_1^2\,\phi_2\,\phi_3 - 2\,\epsilon_2\,\Delta_{12}\,\phi_1\,\phi_3 \Bigr) \\
    & \hspace{39mm} + \epsilon_3 \Bigl( - (1+\epsilon_2)\,\Delta_{23}\,\phi_1^2 - 2\,\Delta_{13} \bigl( (1+\epsilon_2) \,\phi_1 \,\phi_2 - \epsilon_2\,\Delta_{12} \bigr) \Bigr)
    \biggr] \\
    & \hspace{19mm} +  (\Delta_{14})^2\,(1+\epsilon_2)
    \biggl[ (1+\epsilon_3)\,\phi_2^2\,\phi_3^2 + 2\,\epsilon_3 \Bigl(- 2\,\Delta_{23}\,\phi_2\,\phi_3+(\Delta_{23})^2 \Bigr)
    \biggr]
    \\
    & \hspace{19mm} +  (\Delta_{24})^2\,(1+\epsilon_2)
    \biggl[ (1+\epsilon_3)\,\phi_1^2\,\phi_3^2 + 2\,\epsilon_3 \Bigl(- 2\,\Delta_{13}\,\phi_1\,\phi_3+(\Delta_{13})^2 \Bigr)
    \biggr]
    \\
    & \hspace{19mm} +  (\Delta_{34})^2\,(1+\epsilon_3)
    \biggl[ (1+\epsilon_2)\,\phi_1^2\,\phi_2^2 + 2\,\epsilon_2 \Bigl(- 2\,\Delta_{12}\,\phi_1\,\phi_2+(\Delta_{12})^2 \Bigr)
    \biggr]
    \Biggr) \Biggr\} \,.\numberthis
\end{align*}
Using Eqs.~\eqref{eq:hspaceexpvalF4}, \eqref{eq:chispaceexpvalF4}, and \eqref{eq:E1234general}, we can evaluate each of the terms in Eq.~\eqref{eq:F4expvalexpanded} (see Appendix~\ref{app:firstorderdecay}). Note that for any given expectation value, terms with an odd number of $\Delta$s have a relative minus sign to those with an even number, owing to the minus sign before the commutator in Eq.~\eqref{eq:transtrum}. This is reflected in the pre-factor rules in Section~\ref{sec:rules}. } 

Since
\begin{align} \label{eq:thetatotimeordered}
    \Theta_{ij\ldots n} \bra{0^\phi} \phi_i \phi_j \ldots \phi_n \ket{0^\phi} =     \Theta_{ij\ldots n} \bra{0^\phi} \text{T} \{ \phi_i \phi_j \ldots \phi_n \} \ket{0^\phi}\,,
\end{align}
we can use Wick's theorem to express products of fields as Feynman propagators. Each expectation value in Eq.~\eqref{eq:F4reducedE} is shown algebraically in Appendix~\ref{app:firstorderdecay} (unsimplified). To convert these expectation values to probabilities, we simply use Eq.~\eqref{eq:newprob}.

The diagrams can be grouped into different categories based on the topology of the diagram. The diagrams within each category are the same up to the exchange of times. The categories are: 
\begin{itemize}
    \item Disconnected --- One of the initial-state $\chi$s produces a $\phi$ loop. The rest of the diagram is entirely disconnected to this initial state.
    \item Tadpoles --- The $h$ propagator produces a $\phi$ loop.
    \item Oscillations --- The initial-state $\chi$s convert into the $h$ propagator via a loop of two $\phi$ propagators.
    \item Vertex --- The $h$ propagator connects across the $h\phi^2$ vertex, connecting two different $\phi$ `legs' of the diagram.
    \item Self-Energies --- The $h$ propagator connects the same $\phi$ `leg' of the diagram with itself.
\end{itemize}
Disconnected and tadpole diagrams will vanish upon considering correct momentum conservation. Oscillations can be made to have arbitrarily small contributions by varying the mass of the $h$-field compared to the mass of the $\chi$-field. In the analogous QCD process $\gamma \rightarrow q\overline{q} (+g)$, oscillation diagrams would be zero due to colour conservation. 

\begin{figure}[H]
    \centering
    \includegraphics[width=\linewidth]{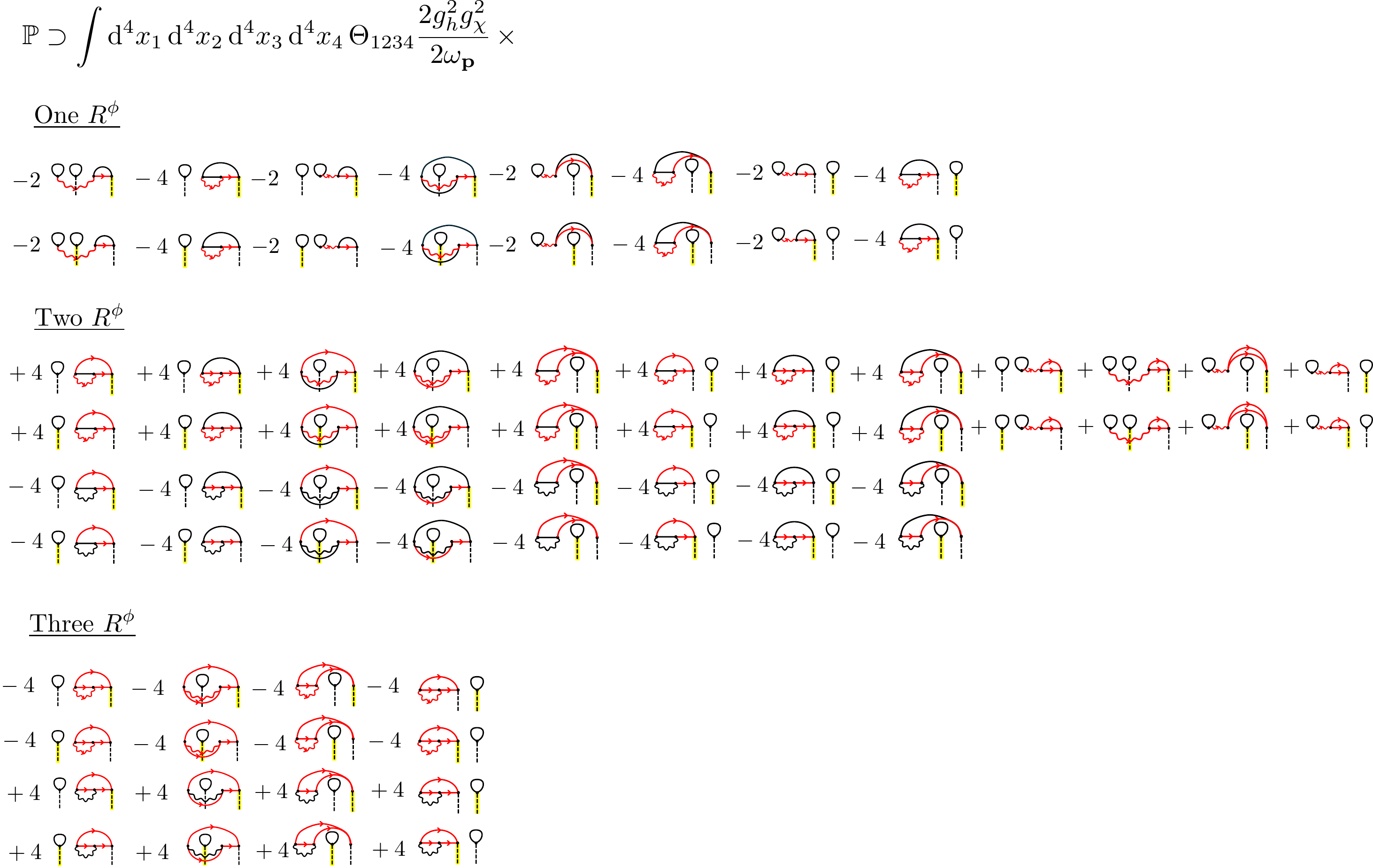}
    \caption{Probability-level diagrammatic representation of \textbf{disconnected} diagrams. The diagrams are sorted by their number of retarded propagators of the $\phi$ field, $R^\phi$.}
    \label{fig:disconnected}
\end{figure}

\begin{figure}[H]
    \centering
    \includegraphics[width=\linewidth]{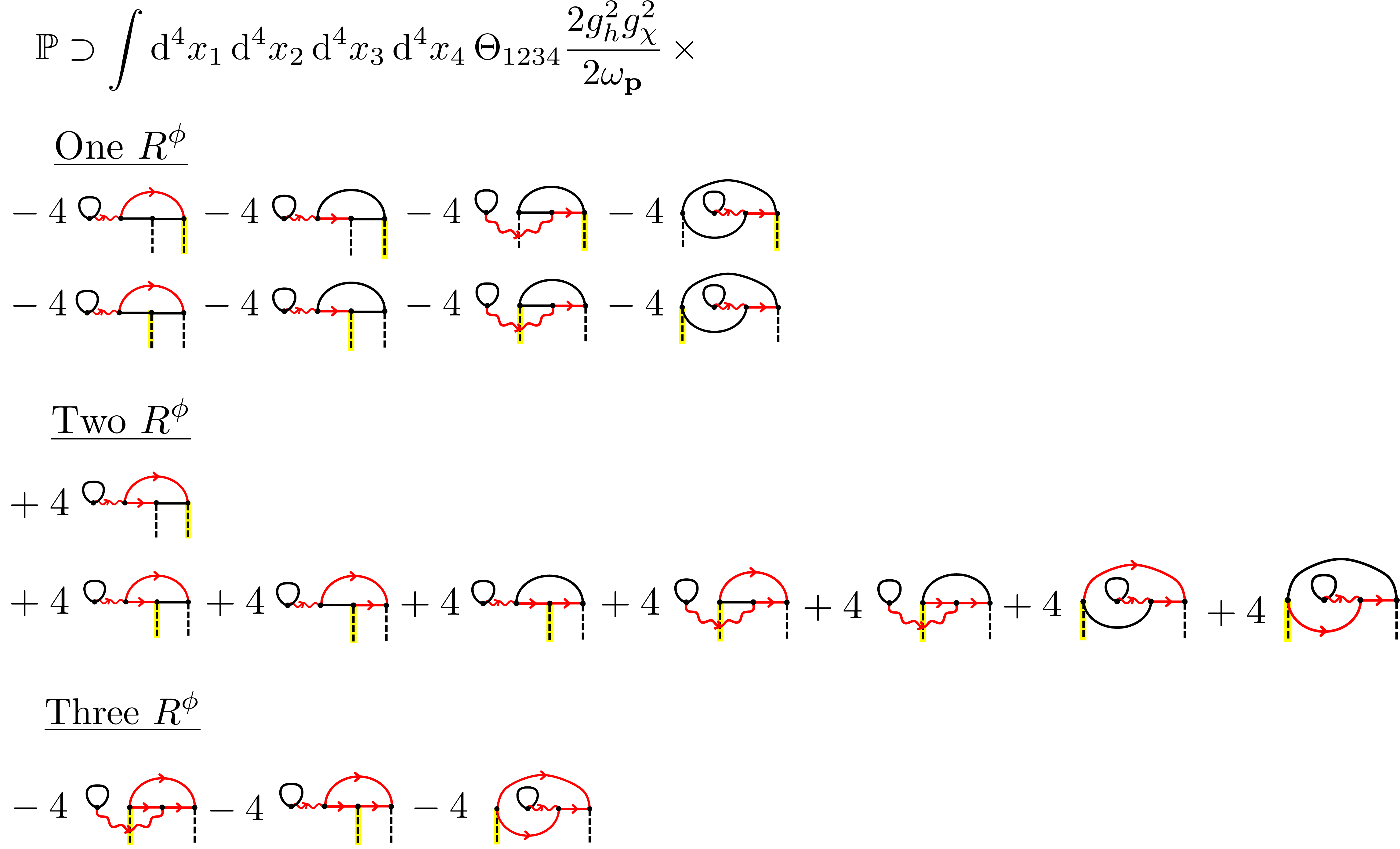}
    \caption{Probability-level diagrammatic representation of \textbf{tadpole} diagrams. The diagrams are sorted by their number of retarded propagators of the $\phi$ field, $R^\phi$.}
    \label{fig:tadpoles}
\end{figure}

\begin{figure}[H]
    \centering
    \includegraphics[width=\linewidth]{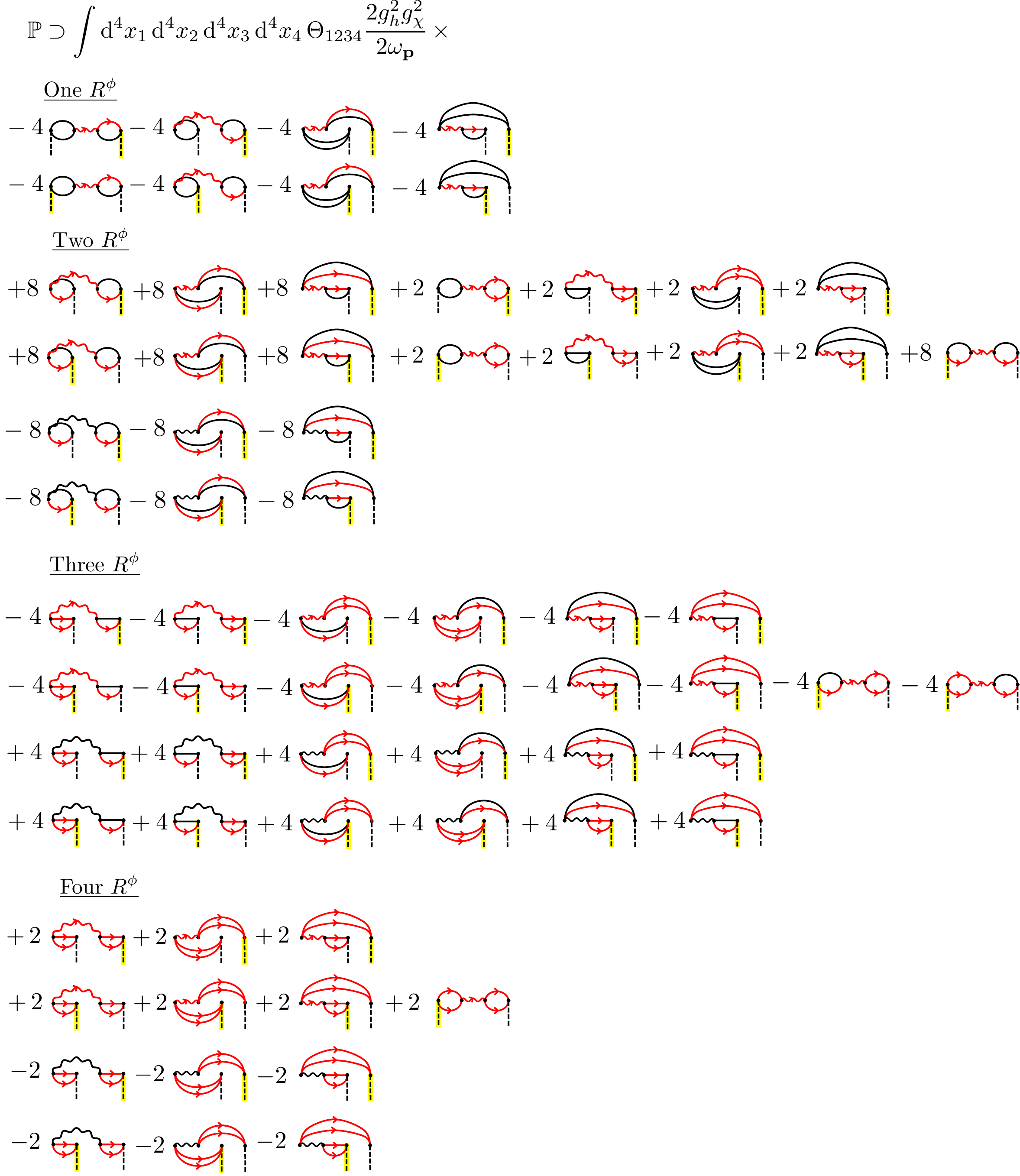}
    \caption{Probability-level diagrammatic representation of \textbf{oscillation} diagrams. The diagrams are sorted by their number of retarded propagators of the $\phi$ field, $R^\phi$.}
    \label{fig:oscillations}
\end{figure}

\begin{figure}[H]
    \centering
    \includegraphics[width=\linewidth]{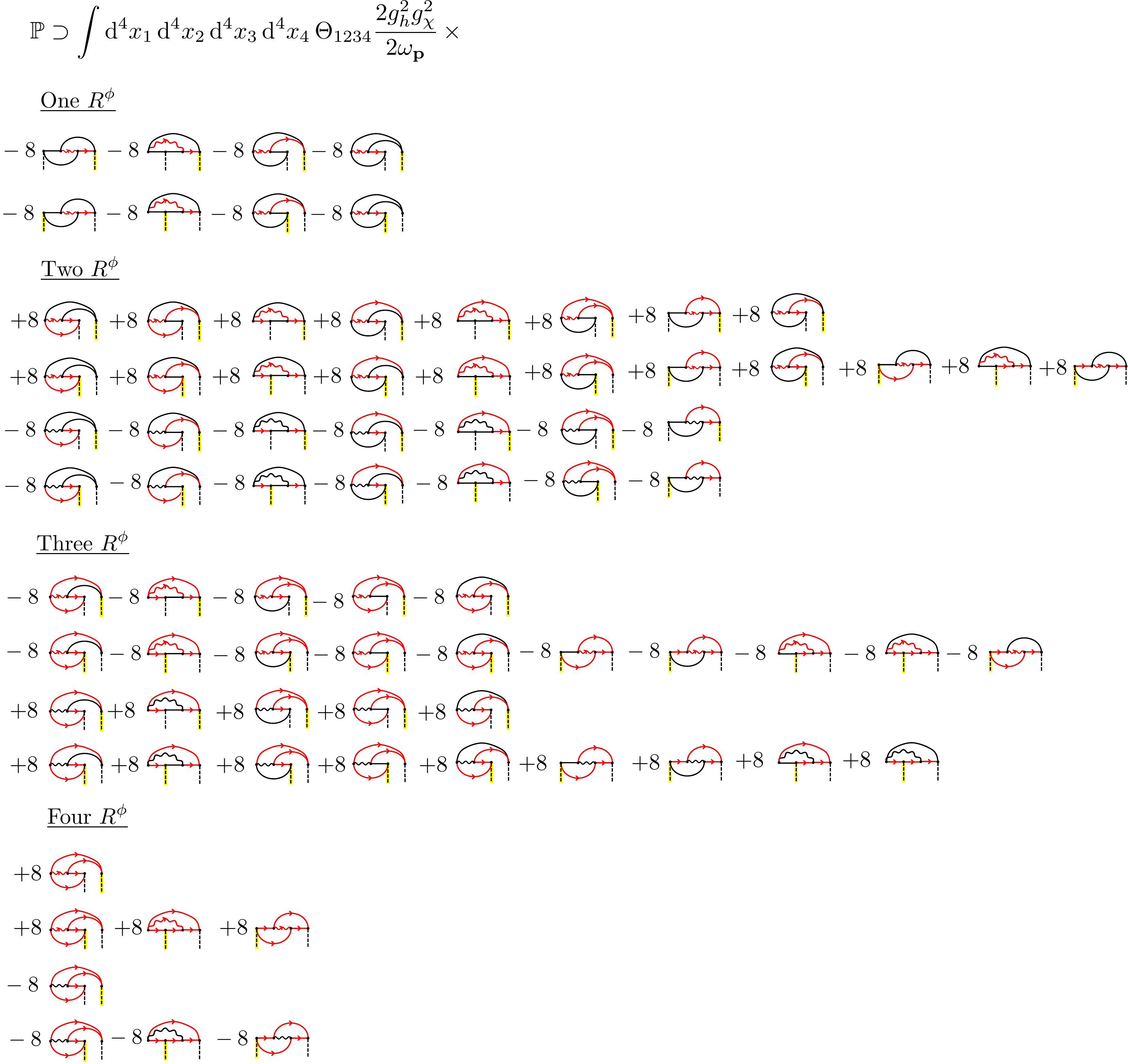}
    \caption{Probability-level diagrammatic representation of \textbf{vertex} diagrams. The diagrams are sorted by their number of retarded propagators of the $\phi$ field, $R^\phi$.}
    \label{fig:vertex}
\end{figure}

\begin{figure}[H]
    \centering
    \includegraphics[width=\linewidth]{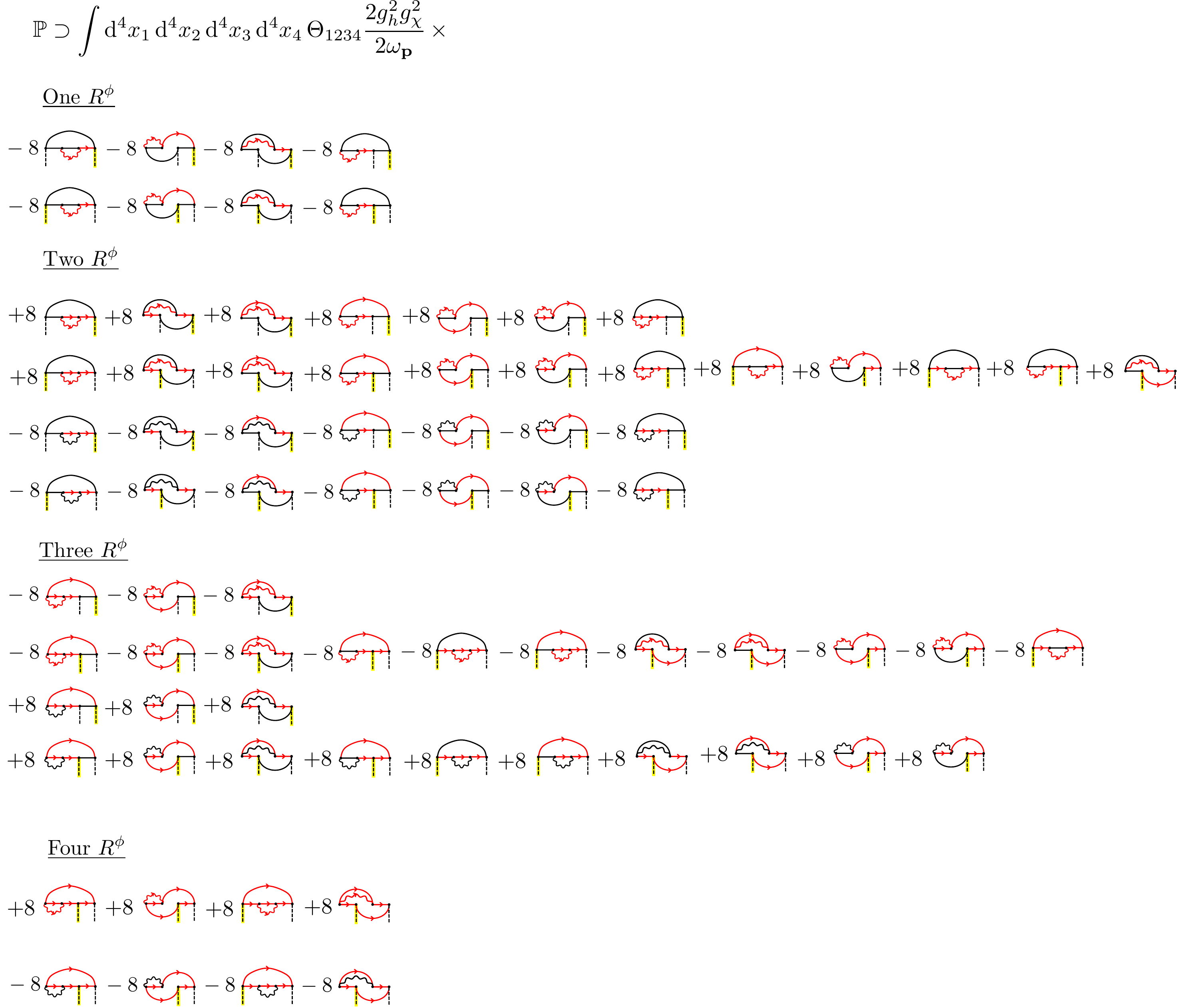}
    \caption{Probability-level diagrammatic representation of \textbf{self-energy} diagrams. The diagrams are sorted by their number of retarded propagators of the $\phi$ field, $R^\phi$.}
    \label{fig:selfs}
\end{figure}

Figs.~\ref{fig:disconnected}--\ref{fig:selfs} show the results diagrammatically. The $h$-field Hadamard function has been rewritten in terms of Feynman and retarded propagators using Eq.~\eqref{eq:Hadamardfunction} and the time-ordering $\Theta$-function. Duplicate diagrams either sum or cancel, reducing the total number of terms. These are exactly the diagrams which would be generated had we used the rules in Section~\ref{sec:rules}. 

Note that there are no distinct `real emission' diagrams, pictured in the centre of Fig.~\ref{fig:feynman decay first}, due to the inclusivity of the effect operator. Instead, these contributions are contained within the `vertex' and `self-energy' diagrams. Since infrared divergences appear in the traditional approach as an artifact of separating vertex, self-energy, and real emission terms, one might be optimistic that the inclusivity of this approach may lead to inherent cancellation of infrared divergences. 

The preceding results could be expressed in terms of Feynman propagators in order to recover more familiar-looking expressions, but doing so would obscure the causal structure. For example, in the next section, this causal approach allows us to go further and see the emergence of the retarded two-point function (Eq.~\eqref{eq:retardedselfenergy}).



\section{Inclusive Annihilation: First Order}\label{sec:annihilation}

We now consider an annihilation process between two particles described by the interaction Hamiltonian
\begin{equation} \label{eq:2to2hamiltonian}
    \mathcal{H}_\text{int} (t_j) = \int \D^3{\mathbf{x}_j} \, \left( g_\psi \psi^2_{j} \chi_{j} + g_\phi \phi^2_{j} \chi_{j} \right)\,,
\end{equation}
where $\psi_j\equiv\psi(x_j)$ is another real scalar field.
Consider the initial density matrix
\begin{equation}
    \rho_0 = \ket{0^\chi\,p_1^\psi p_2^\psi\,0^\phi} \bra{{0^\chi\,p_1^\psi p_2^\psi\,0^\phi}} \,, \label{eq:chipsiphi:initial}
\end{equation}
i.e., the system initially has two $\psi$ particles, of momentum $p_1$ and $p_2$, and no other field excitations. We choose the effect operator
\begin{equation}
\begin{split}
    E = \sum_{n,\alpha} \ket{n^\chi\,0^\psi\,\alpha^\phi} \bra{n^\chi\,0^\psi\,\alpha^\phi} = \mathbb{I}^\chi \, \ket{0^\psi} \bra{0^\psi} \, \mathbb{I}^\phi \; , \\
    \text{i.e., } \quad E^\chi = \mathbb{I}^\chi \; , \quad E^\psi = \ket{0^\psi} \bra{0^\psi} \; , \quad E^{\phi} = \mathbb{I}^\phi \; ,
    \label{Escattering2to2}
\end{split}
\end{equation}
such that we consider all possible final states with zero quanta of the $\psi$-field. 

Since the structure of the Hamiltonian differs slightly to the decay studied in Sections~\ref{sec:decay_zeroth} and \ref{sec:firstorderdecay}, our notation is now
\begin{alignat}{3}
    \text{Hilbert space $\chi$: }& \hspace{7mm} &&\mathcal{E}^{\dots \psi}_{\dots k} \coloneqq \, \frac{1}{i} \comm{\mathcal{E}^{\dots}_{\dots}}{g_\psi \chi _k}, \hspace{7mm} &&\mathcal{E}^{\dots \psi}_{\dots \ul{k}} \coloneqq \,\acomm{\mathcal{E}^{\dots}_{\dots}}{g_\psi \chi_k}, \nonumber \\
    & \hspace{7mm} &&\mathcal{E}^{\dots \phi }_{\dots k} \coloneqq \,\frac{1}{i} \comm{\mathcal{E}^{\dots}_{\dots}}{g_\phi  \chi_k}, \hspace{7mm} &&\mathcal{E}^{\dots \phi }_{\dots \ul{k}} \coloneqq \,\acomm{\mathcal{E}^{\dots}_{\dots}}{g_\phi  \chi_k}, \label{eq:curlyEdots2to2} \\
    \text{Hilbert space $\psi$: }& \hspace{7mm} &&E^\psi_{\dots k} \coloneqq \frac{1}{i} \comm{E^\psi_{\dots}}{\psi^2_k}, \hspace{10mm} &&E^\psi_{\dots \ul{k}} \coloneqq \acomm{E^\psi_{\dots}}{\psi^2_k}, \nonumber \\
    \text{Hilbert space $\phi$: }& \hspace{7mm} &&E^\phi _{\dots k} \coloneqq \frac{1}{i} \comm{E^\chi _{\dots}}{\phi^2 _k}, \hspace{10mm} &&E^\phi _{\dots \ul{k}} \coloneqq \acomm{E^\phi _{\dots}}{\phi^2 _k}, \label{eq:Edots2to2}
\end{alignat}
and $\mathcal{E} = E^\chi$.

The lowest-order contributions come from the process $\psi\psi \rightarrow \chi$, contained within $\mathcal{F}_2$, which are trivial and uninteresting. Instead, we consider the next order of contributions, which come from $\mathcal{F}_4$. Specifically, we isolate and examine the $\order{g_\psi^2 g_\phi^2}$ contributions, which correspond to the $a=2$ term in $\mathcal{F}_4$,
\begin{align}
    \mathcal{F}^{(a=2)}_4 &= \frac{1}{16} \int \D^3 {\mathbf{x}_{1}} \,\D^3 {\mathbf{x}_{2}} \,\D^3 {\mathbf{x}_{3}} \,\D^3 {\mathbf{x}_{4}} E^\psi_{(\uc{1}\uc{2}}\,E^\phi_{\uc{3}\uc{4})}\,\mathcal{E}^{(\psi \psi \phi \phi )}_{(\ub{1}\ub{2} \ub{3}\ub{4})} ~.
\label{eq:F4a=2for2to2}
\end{align}
Similar to the first-order $h$ corrections to the inclusive decay in Section~\ref{sec:firstorderdecay}, the inclusive effect operators $\mathcal{E} = \mathbb{I}^\chi$ and $E^\phi = \mathbb{I}^\phi$ result in the only non-zero terms in Eq.~\eqref{eq:F4a=2for2to2} having the general form
\begin{align}
    E^\psi_{1\uc{j}}\,E^\phi_{\ul{k}\uc{l}}\,\mathcal{E}^{\psi (\psi \phi \phi )}_{\ul{1}\ub{2} \ub{3}\ub{4}} ~.
\label{eq:generalE2to2}
\end{align}
As in Section~\ref{sec:firstorderdecay}, there will always be four non-underlined indices. Due to Eqs.~\eqref{eq:curlyEdots2to2} and \eqref{eq:Edots2to2}, this will always result in a factor of $(1/i)^4=1$, so these factors of $i$ will be ignored for the rest of this section.

\begin{figure}
    \centering
    \includegraphics[width=\linewidth]{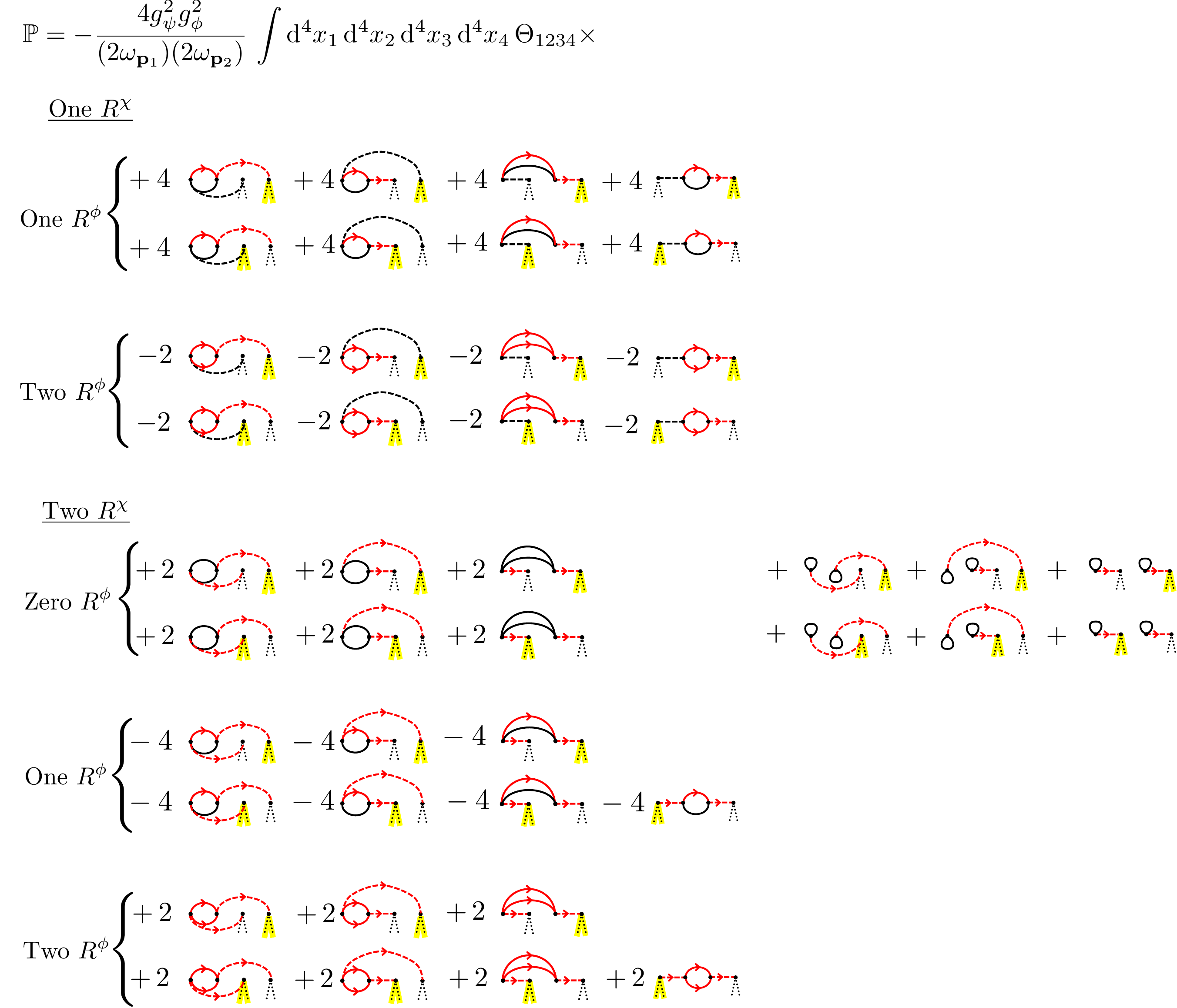}
    \caption{Probability-level diagrammatic representation of the annihilation process expressed in Eq.~\eqref{eq:expval2to2}. Terms are sorted by the number of retarded propagators for each field, $R^\chi$ and $R^\phi$.}
    \label{fig:2to2sorted}
\end{figure}

The expectation value of the $\psi$-space operator is
\begin{align}
    \bra{p_1^\psi p_2^\psi} E^\psi_{1\uc{j}} \ket{p_1^\psi p_2^\psi} =& \bra{p_1^\psi p_2^\psi}\biggl[\Bigl[\ket{0^\psi}\bra{0^\psi},\psi_1^2\Bigr],\psi_j^2\biggr]_{\eta_j} \ket{p_1^\psi p_2^\psi}\nonumber\\
    =& \frac{4}{ (2\omega_{p_1})(2\omega_{p_2})} \bigl( -e^{ip_1\cdot x_1}e^{ip_2\cdot x_1}e^{-ip_1\cdot x_j}e^{-ip_2\cdot x_j} +\eta_j e^{ip_1\cdot x_j}e^{ip_2\cdot x_j}e^{-ip_1\cdot x_1}e^{-ip_2\cdot x_1} \bigr) \,.
\end{align}
The expectation value of the $\phi$-space operator is
\begin{align}
    \bra{0^\phi} E^\phi_{\ul{k}\uc{l}} \ket{0^\phi} &= \bra{0^\phi}\biggl[\Bigl\{\mathbb{I}^\phi,\phi_k^2\Bigr\},\phi_l^2\biggr]_{\lambda_l} \ket{0^\phi} = 2 \bra{0^\phi} \bigl[ \phi_k^2 , \phi_l^2 \bigr]_{\lambda_l} \ket{0^\phi}\nonumber\\
    &= 2\,(1+\lambda_j) \bra{0^\phi} \phi_k^2 \phi_l^2 \ket{0^\phi} - 8\, \lambda_l\, \Delta^\phi_{kl} \bra{0^\phi}\phi_k \phi_l \ket{0^\phi} + 4\,\lambda_l \,(\Delta^\phi_{kl})^2 \,.
\end{align}
The $\chi$-space operator is
\begin{align} \label{eq:2to2chieffect}
    \mathcal{E}_{\ul{1}\ub{2}\ub{3}\ub{4}} &= g_\psi^2 g_\phi^2\Biggl[ \biggl[ \Bigl[ \bigl\{ \mathbb{I}^\chi ,\chi_1 \bigr\} , \chi_2 \Bigr]_{\epsilon_2} , \chi_3 \biggr]_{\epsilon_3} , \chi_4 \Biggr]_{\epsilon_4}
    \nonumber \\
    &= g_\psi^2 g_\phi^2\,(1+\epsilon_4) \biggl( 2\,(1+\epsilon_3) \Bigl( (1+\epsilon_2) \chi_1 \chi_2 \chi_3 \chi_4 - \epsilon_2 \Delta^\chi_{12} \chi_3 \chi_4 \Bigr) \nonumber\\
    &\qquad \qquad \qquad \qquad - 2\,(1+\epsilon_2)\,\epsilon_3 \Bigl( \Delta^\chi_{13} \chi_2 \chi_4 + \Delta^\chi_{23} \chi_1 \chi_4 \Bigr) \biggr) \nonumber\\
    & \qquad -2\,g_\psi^2 g_\phi^2\,\epsilon_4 \biggl( (1+\epsilon_2) \Delta^\chi_{14} \Bigl( (1+\epsilon_3) \chi_2 \chi_3 - \epsilon_3 \Delta^\chi_{23} \Bigr) \nonumber\\
    & \qquad \qquad \qquad \qquad+ (1+\epsilon_2) \Delta^\chi_{24} \Bigl( (1+\epsilon_3) \chi_1 \chi_3 - \epsilon_3 \Delta^\chi_{13} \Bigr) \nonumber\\
    & \qquad \qquad \qquad \qquad + (1+\epsilon_3) \Delta^\chi_{34} \Bigl( (1+\epsilon_2)\chi_1 \chi_2 - \epsilon_2 \Delta^\chi_{12} \Bigr) \biggr)\,.
\end{align}
The sum of operators over all non-zero permutations of time indices is
\begin{align}
    E^\psi_{1(\uc{2}}\,E^\phi_{\ul{3}\uc{4})}\,\mathcal{E}^{\psi (\psi \phi \phi )}_{\ul{1}\ub{2} \ub{3}\ub{4}} &= 
    E^\psi_{1 2}\,E^\phi_{\ul{3} 4}\,\mathcal{E}^{\psi \psi \phi \phi}_{\ul{1}\ul{2} 3\ul{4}} + 
    E^\psi_{1 \ul{2}}\,E^\phi_{\ul{3} 4}\,\mathcal{E}^{\psi \psi \phi \phi}_{\ul{1}2 3\ul{4}} + 
    E^\psi_{1 2}\,E^\phi_{\ul{3} \ul{4}}\,\mathcal{E}^{\psi \psi \phi \phi}_{\ul{1}\ul{2} 34} +
    E^\psi_{1 \ul{2}}\,E^\phi_{\ul{3} \ul{4}}\,\mathcal{E}^{\psi \psi \phi \phi}_{\ul{1}234} \nonumber\\
    & +E^\psi_{1 3}\,E^\phi_{\ul{2} 4}\,\mathcal{E}^{\psi \phi \psi  \phi}_{\ul{1}2\ul{3}\ul{4}} + 
    E^\psi_{1 \ul{3}}\,E^\phi_{\ul{2} 4}\,\mathcal{E}^{\psi \phi \psi \phi}_{\ul{1}2 3\ul{4}} + 
    E^\psi_{1 3}\,E^\phi_{\ul{2} \ul{4}}\,\mathcal{E}^{\psi \phi \psi \phi}_{\ul{1}2\ul{3}4} +
    E^\psi_{1 \ul{3}}\,E^\phi_{\ul{2} \ul{4}}\,\mathcal{E}^{\psi  \phi \psi \phi}_{\ul{1}234} \nonumber\\
    &+E^\psi_{1 4}\,E^\phi_{\ul{2} 3}\,\mathcal{E}^{\psi \phi \phi  \psi}_{\ul{1}2\ul{3}\ul{4}} + 
    E^\psi_{1 \ul{4}}\,E^\phi_{\ul{2} 3}\,\mathcal{E}^{\psi \phi \phi \psi}_{\ul{1}2 \ul{3}4} + 
    E^\psi_{1 4}\,E^\phi_{\ul{2} \ul{3}}\,\mathcal{E}^{\psi \phi \phi \psi}_{\ul{1}23\ul{4}} +
    E^\psi_{1 \ul{4}}\,E^\phi_{\ul{2} \ul{3}}\,\mathcal{E}^{\psi  \phi \phi \psi}_{\ul{1}234} ~.
\label{eq:Esum2to2}
\end{align}
From Eq.~\eqref{eq:2to2chieffect}, we have
\begin{align}
    & \mathcal{E}^{(\psi \psi \phi \phi)}_{\ul{1}2 3\ul{4}} =
    \mathcal{E}^{(\psi \psi \phi \phi)}_{\ul{1}234} = 0 ~,
\end{align}
so Eq.~\eqref{eq:Esum2to2} becomes
\begin{align}
    E^\psi_{1(\uc{2}}\,E^\phi_{\ul{3}\uc{4})}\,\mathcal{E}^{\psi (\psi \phi \phi )}_{\ul{1}\ub{2} \ub{3}\ub{4}} &= 
    E^\psi_{1 2}\,E^\phi_{\ul{3} 4}\,\mathcal{E}^{\psi \psi \phi \phi}_{\ul{1}\ul{2} 3\ul{4}} + 
    E^\psi_{1 2}\,E^\phi_{\ul{3} \ul{4}}\,\mathcal{E}^{\psi \psi \phi \phi}_{\ul{1}\ul{2} 34} + E^\psi_{1 3}\,E^\phi_{\ul{2} 4}\,\mathcal{E}^{\psi \phi \psi  \phi}_{\ul{1}2\ul{3}\ul{4}}  + 
    E^\psi_{1 3}\,E^\phi_{\ul{2} \ul{4}}\,\mathcal{E}^{\psi \phi \psi \phi}_{\ul{1}2\ul{3}4}  \nonumber\\
    &+ E^\psi_{1 4}\,E^\phi_{\ul{2} 3}\,\mathcal{E}^{\psi \phi \phi  \psi}_{\ul{1}2\ul{3}\ul{4}} + 
    E^\psi_{1 \ul{4}}\,E^\phi_{\ul{2} 3}\,\mathcal{E}^{\psi \phi \phi \psi}_{\ul{1}2 \ul{3}4} ~.
\label{eq:Esum2to2simplified}
\end{align}
Using our equations for each Hilbert space and substituting into Eq.~\eqref{eq:newprob}, we find
{\allowdisplaybreaks
\begin{align}
    \mathbb{P} =& -\frac{4 \,g_\psi^2 g_\phi^2}{   (2\omega_{p_1})(2\omega_{p_2})} \int \D^4 {x_{1}} \,\D^4 {x_{2}} \,\D^4 {x_{3}} \,\D^4 {x_{4}} \,\Theta_{1234} \nonumber \\
    & \Biggl( \Bigl( e^{ip_1\cdot x_1}e^{ip_2\cdot x_1}e^{-ip_1\cdot x_2}e^{-ip_2\cdot x_2} + e^{ip_1\cdot x_2}e^{ip_2\cdot x_2}e^{-ip_1\cdot x_1}e^{-ip_2\cdot x_1} \Bigr) \nonumber\\
    & \Bigl( 2\Delta^\phi_{34} \expval{\phi_3 \phi_4} - (\Delta^\phi_{34})^2 \Bigr) \nonumber\\
    & \Bigl(
    2\,\Delta^\chi_{13} \,\expval{\chi_2 \,\chi_4} + 2\,\Delta^\chi_{23} \,\expval{\chi_1 \,\chi_4} - \Delta^\chi_{14} \Delta^\chi_{23} - \Delta^\chi_{13} \Delta^\chi_{24} \Bigr) \nonumber\\
    & + \Bigl( e^{ip_1\cdot x_1}e^{ip_2\cdot x_1}e^{-ip_1\cdot x_2}e^{-ip_2\cdot x_2} +e^{ip_1\cdot x_2}e^{ip_2\cdot x_2}e^{-ip_1\cdot x_1}e^{-ip_2\cdot x_1} \Bigr) \nonumber\\
    & \Bigl( \expval{ \phi_3^2 \phi_4^2 } -2\Delta^\phi_{34} \expval{\phi_3 \phi_4} + (\Delta^\phi_{34})^2 \Bigr) 
    \Bigl( \Delta^\chi_{14} \Delta^\chi_{23} + \Delta^\chi_{13} \Delta^\chi_{24} \Bigr) \nonumber\\
    & + \Bigl( e^{ip_1\cdot x_1}e^{ip_2\cdot x_1}e^{-ip_1\cdot x_3}e^{-ip_2\cdot x_3} + e^{ip_1\cdot x_3}e^{ip_2\cdot x_3}e^{-ip_1\cdot x_1}e^{-ip_2\cdot x_1} \Bigr) \nonumber\\
    & \Bigl( 2\Delta^\phi_{24} \expval{\phi_2 \phi_4} - (\Delta^\phi_{24})^2 \Bigr)
    \Bigl(
    2\,\Delta^\chi_{12} \,\expval{\chi_3 \,\chi_4} -\Delta^\chi_{12}\Delta^\chi_{34} \Bigr) \nonumber\\
    & + \Bigl( e^{ip_1\cdot x_1}e^{ip_2\cdot x_1}e^{-ip_1\cdot x_3}e^{-ip_2\cdot x_3} + e^{ip_1\cdot x_3}e^{ip_2\cdot x_3}e^{-ip_1\cdot x_1}e^{-ip_2\cdot x_1} \Bigr) \nonumber\\
    & \Bigl( \expval{ \phi_2^2 \phi_4^2 } -2\Delta^\phi_{24} \expval{\phi_2 \phi_4} + (\Delta^\phi_{24})^2 \Bigr)
    \Bigl(
    \Delta^\chi_{12}\Delta^\chi_{34} \Bigr) \nonumber\\
    & + \Bigl( e^{ip_1\cdot x_1}e^{ip_2\cdot x_1}e^{-ip_1\cdot x_4}e^{-ip_2\cdot x_4} + e^{ip_1\cdot x_4}e^{ip_2\cdot x_4}e^{-ip_1\cdot x_1}e^{-ip_2\cdot x_1} \Bigr) \nonumber\\
    & \Bigl( 2\Delta^\phi_{23} \expval{\phi_2 \phi_3 } - (\Delta^\phi_{23})^2 \Bigr)
    \Bigl(
    2\Delta^\chi_{12} \, \expval{\chi_3 \, \chi_4} - \Delta^\chi_{12}\Delta^\chi_{34} \Bigr) \nonumber\\
    & + \Bigl( e^{ip_1\cdot x_1}e^{ip_2\cdot x_1}e^{-ip_1\cdot x_4}e^{-ip_2\cdot x_4} - e^{ip_1\cdot x_4}e^{ip_2\cdot x_4}e^{-ip_1\cdot x_1}e^{-ip_2\cdot x_1} \Bigr) \nonumber\\
    & \Bigl( 2\Delta^\phi_{23} \expval{\phi_2 \phi_3 } - (\Delta^\phi_{23})^2 \Bigr)
    \Bigl(
     \Delta^\chi_{12}\Delta^\chi_{34} \Bigr) \Biggr)
    ~,
\label{eq:expval2to2}
\end{align}
where $\expval{\ldots}$ is shorthand for $\bra{0^\phi} \ldots \ket{0^\phi}$ or $\bra{0^\chi} \ldots \ket{0^\chi}$. Due to the appearance of $\Theta_{1234}$ (from Eq.~\eqref{eq:newprob}), we can interpret the product of fields as the time-ordered product, and thus use Wick's theorem (as in Eq.~\eqref{eq:thetatotimeordered}). This results in Feynman propagators. Eq.~\eqref{eq:expval2to2} is shown diagrammatically in Fig.~\ref{fig:2to2sorted}, after simplification. Again, the diagrams are exactly those which would be generated using the rules in Section~\ref{sec:rules}. }

\begin{figure}
    \centering
    \includegraphics[width=0.4\linewidth]{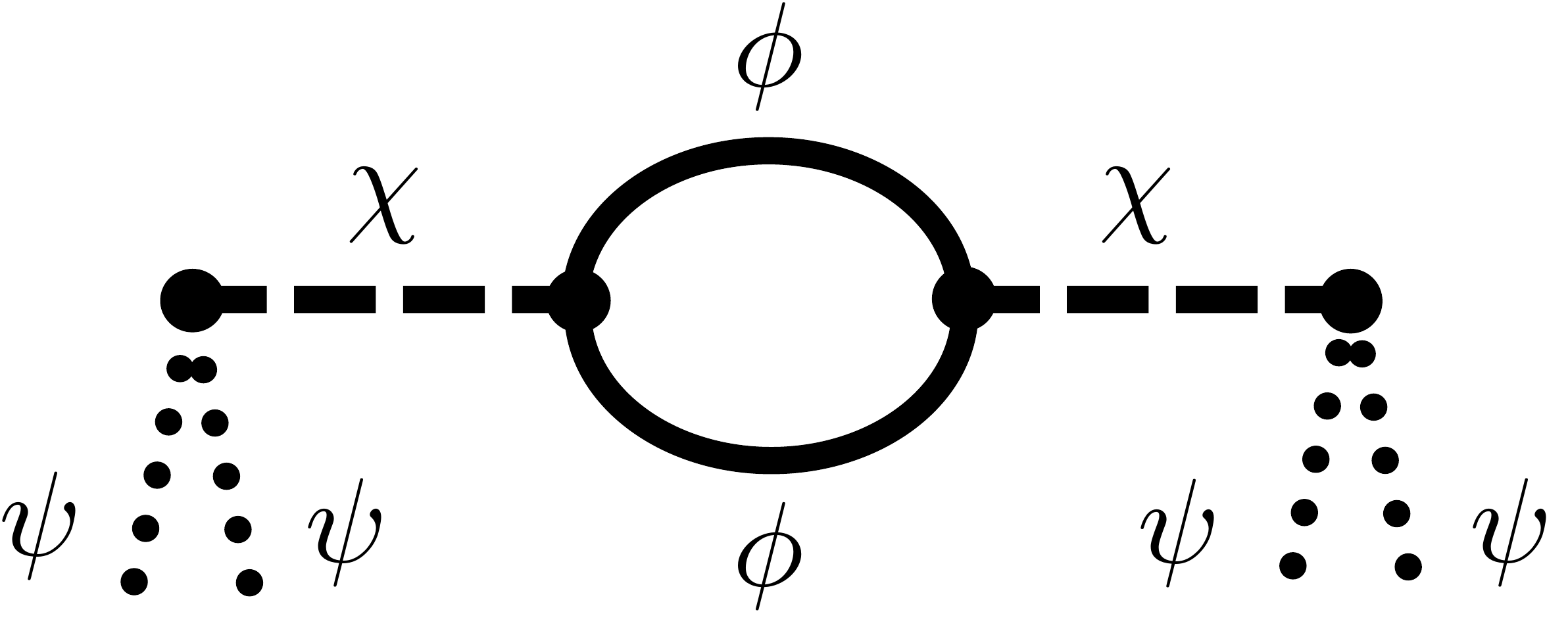}
    \caption{The topology of all of the diagrams in Fig.~\ref{fig:2to2sorted}.}
    \label{fig:2to2 topology}
\end{figure}

\begin{figure}
    \centering
    \includegraphics[width=\linewidth]{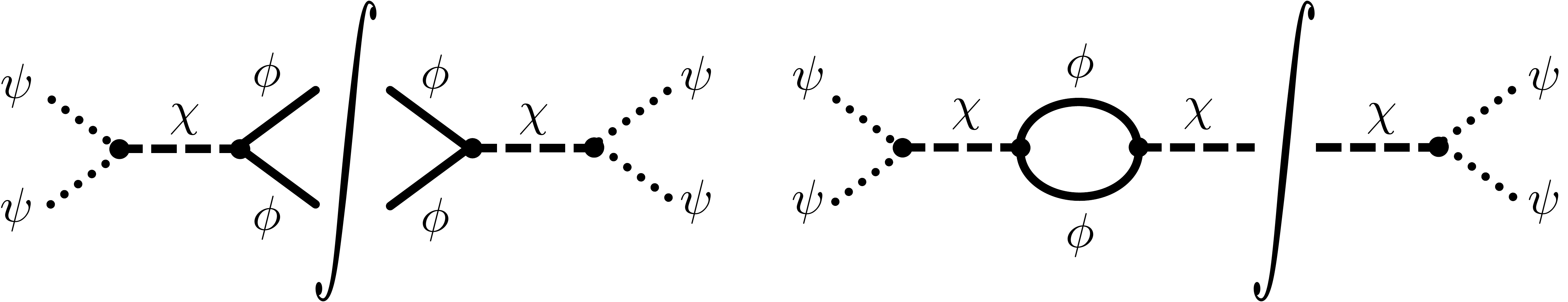}
    \caption{Feynman diagrams (and conjugate diagrams) for the two annihilation processes. The curved line separates an amplitude from a conjugate amplitude. These are the annihilation processes which are encoded in the diagrams in Fig.~\ref{fig:2to2sorted}. \textbf{Left:} $\psi\psi \rightarrow \chi \rightarrow \phi\phi$. \textbf{Right:} $\psi\psi\rightarrow\chi$ with a $\phi$ self-energy loop. The conjugate of this diagram also contributes.}
    \label{fig:annihilation processes}
\end{figure}

The disconnected diagrams can be ignored since they are not kinematically allowed for positive, non-zero initial momenta. All of the remaining diagrams have the topology shown in Fig.~\ref{fig:2to2 topology}, differing by their time-orderings and combinations of Feynman and retarded propagators. Since our approach is inclusive over final states, these diagrams include two annihilation processes: the 2-to-2 process $\psi\psi \rightarrow \chi \rightarrow \phi\phi$ and the 2-to-1 process $\psi\psi\rightarrow\chi$ with a $\phi$ self-energy loop. The Feynman diagrams for these processes are shown in Fig.~\ref{fig:annihilation processes}. Note that the diagrams in Fig.~\ref{fig:2to2sorted} do not neatly separate into 2-to-2 and 2-to-1 diagrams, highlighting the intrinsic inclusivity of the result.

The diagrams in Fig.~\ref{fig:2to2sorted} can be simplified by introducing the retarded self-energy~\cite{VANEIJCK1992305, PhysRevD.43.1269},
\begin{equation}\label{eq:retardedselfenergy}
    \Pi^{\text{R}}_{xy} = \frac{g_\phi^2}{2} \left[ 2\,F^\phi_{xy}\,R^\phi_{xy} - \left( R^\phi_{xy} \right)^2 \right]\,,
\end{equation}
reflecting the manifest causality of this approach. This simplification suggests the existence of more fundamental rules than those given in Section~\ref{sec:rules}, i.e., rules involving manifestly causal objects such as the retarded self-energy.

The diagrams can be simplified further by realising that all possible time orderings are present, since the retarded propagators vanish when the arguments are not time-ordered (see Eq.~\eqref{eq:RetardedDefinition}). Consequently, the time-ordered integral can be replaced by an integral over all times, and we need only keep one of each unique diagram. This process is shown in Fig.~\ref{fig:timesymmetry}, and is proven analytically in Appendix~\ref{app:timesymmetry} for the diagrams with a loop of Feynman propagators (a similar procedure has been used for all diagrams).
\begin{figure}
    \centering
    \includegraphics[width=\linewidth]{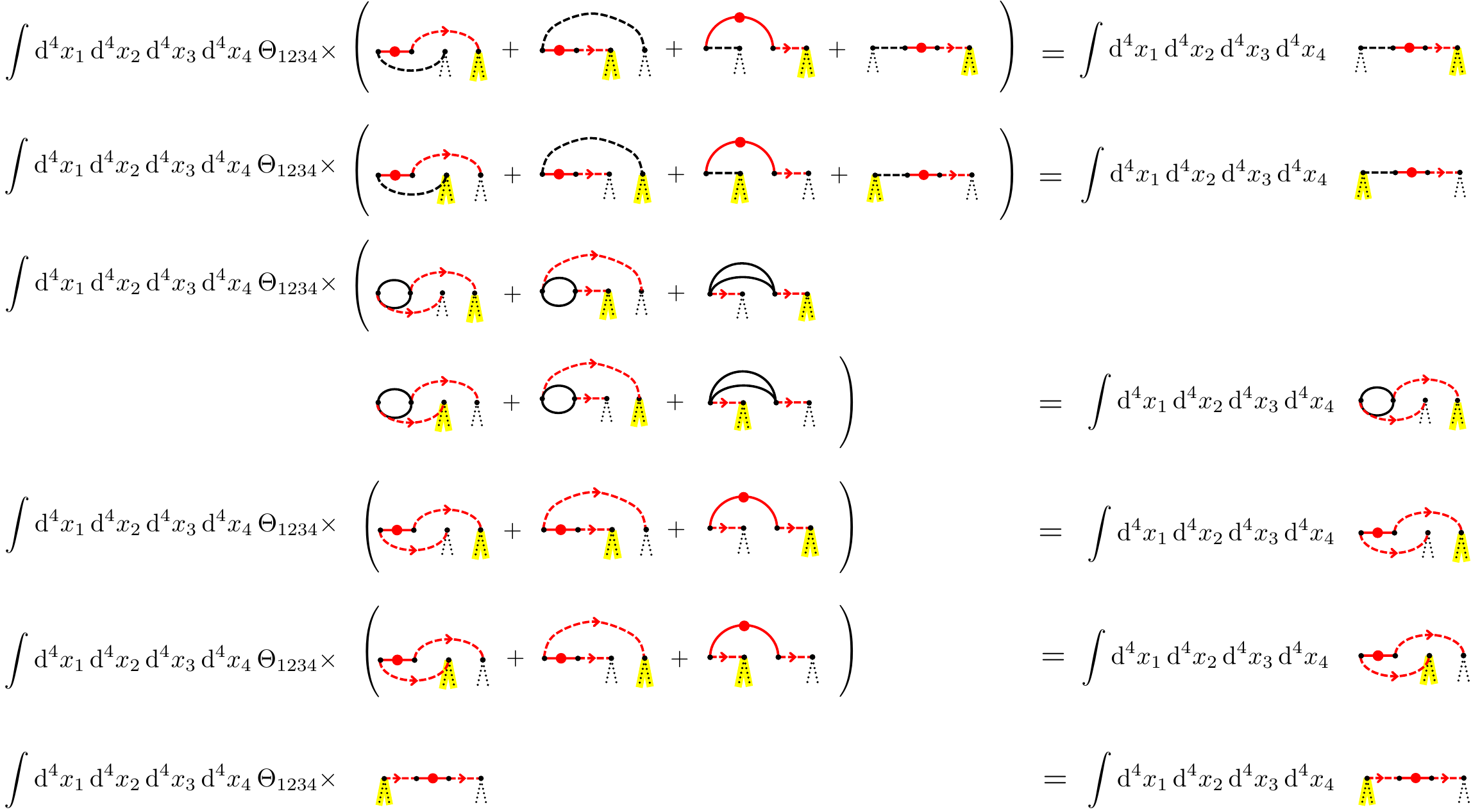}
    \caption{Simplification of the annihilation diagrams. Note the absence of time-ordering on the right-hand side. The retarded self-energy, given by Eq.~\eqref{eq:retardedselfenergy}, is denoted by a red line with a solid red circle in the middle.}
    \label{fig:timesymmetry}
\end{figure}

\begin{figure}
    \centering
    \includegraphics[width=\linewidth]{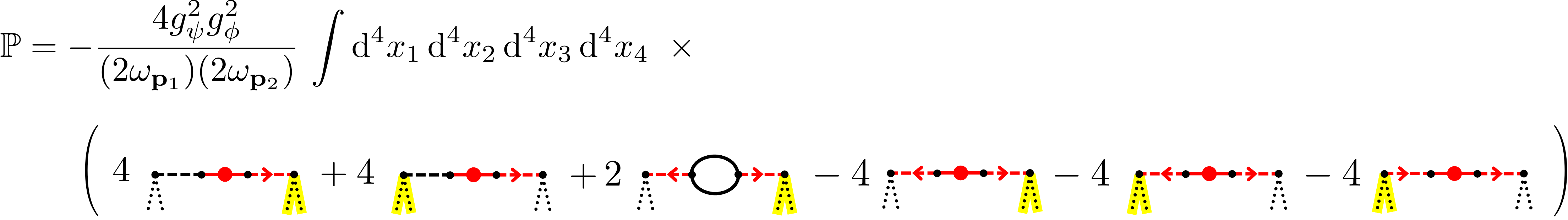}
    \caption{The simplified result for the fully inclusive annihilation probability. The retarded self-energy, given by Eq.~\eqref{eq:retardedselfenergy}, is denoted by a red line with a solid red circle in the middle.}
    \label{fig:finalannihilation}
\end{figure}

The diagrams resulting from these simplifications are shown in Fig.~\ref{fig:finalannihilation}. Further simplification using propagator identities is possible, but we choose to keep the result solely in terms of Feynman and retarded propagators to highlight the causal structure.


\section{Summary}\label{sec:summary}

We have applied a novel, probability-level QFT formalism to scalar field scattering processes in which causality is manifest. In scalar field theory, causality is encoded in the commutator of fields, which appear as a result of applying the Baker-Campbell-Hausdorff lemma to the transition probability. This formalism results in new probability-level diagrams, and we have conjectured the rules to generate the complete set of all diagrams for a scattering process which is fully inclusive over final states that do not contain initial-state particles. We have used the algebraic formalism to calculate the total probabilities for particle decay and the annihilation of two particles, both at fixed order. These results align with those expected from a traditional calculation and corroborate the general diagrammatic rules.

Since the diagrams correspond to the probability directly and involve retarded propagators, causality is manifest. The appearance of other causal structures, such as the retarded self-energy, suggests the existence of a more fundamental set of rules in terms of these causal objects. In particular, there may be a link between these rules and the Kobes-Semenoff unitary cutting rules~\cite{KOBES1985714, KOBES1986329} in a similar fashion to the discussion in Ref.~\cite{Dickinson:2013lsa}. 

The diagrammatic method developed in this paper will generalise to gauge theories such as QED and QCD. In these theories, individual contributions are infrared-divergent but finite once regularised and combined. Our probability-level method results in retarded propagators appearing in loops, and `real emission' contributions are accounted for in the `self-energy' and `vertex' terms. We hope that these features may help avoid the infrared divergences in gauge theories by satisfying the KLN theorem implicitly. In this context, it would be of interest to understand how the present work relates to manifestly causal Loop-Tree Duality~\cite{capatti2020manifestly, GermanRodrigo_2, GermanRodrigo_1, LTD:2024yrb}, Local Unitarity~\cite{Capatti:2020xjc, Capatti:2021bsm, Capatti2022}, and the idea of local infrared safety~\cite{Sterman:2023xdj}.

Future work should include understanding how the rules are adapted for semi-inclusive effect operators, such as those discussed in Ref.~\cite{Dickinson:2017uit}. Moreover, it remains to establish how manifestly causal probabilities can be extracted from causal $n$-point functions by means of an LSZ-like reduction procedure.

\begin{acknowledgments}

This work was supported by the Science and Technology Facilities Council (STFC) [Grant No.~ST/X00077X/1], a United Kingdom Research and Innovation (UKRI) Future Leaders Fellowship [Grant Nos.~MR/V021974/1 and~MR/V021974/2], a Nottingham Research Fellowship from the University of Nottingham, and a Leverhulme Trust Research Leadership Award [Grant No.~RL-2016-028]. RJ would like to thank Callsign Ltd for their financial support. 

\end{acknowledgments}

\section*{Data Access Statement}

No data were created or analysed in this study.

\appendix

\section{Propagator Definitions}\label{app:propagators}

Throughout this paper, we use the following definitions and notation:

{ \allowdisplaybreaks
\begin{align}
\text{Positive Wightman Function: }  &\notag\\[0.5em]
 \Delta_{xy}^{\phi(>)} \equiv \Delta^{\phi(>)} (x-y) &= \expval**{\phi (x) \phi (y)}{0^\phi} \label{eq:Wightmanfunction} \\
&= \int \frac{\dd[3]{\mathbf{k}}}{(2\pi)^3} \frac{1}{2 \omega_k} e^{-ik \cdot (x - y)} \Big|_{k_0 = \omega_k} ~, \notag \\[1em]
\text{Negative Wightman Function: }  &\notag\\[0.5em]
 \Delta_{xy}^{\phi(<)} \equiv \Delta^{\phi(<)} (x-y) &= \expval**{\phi (y) \phi (x)}{0^\phi} \label{eq:NegativeWightmanfunction} \\
&= \int \frac{\dd[3]{\mathbf{k}}}{(2\pi)^3} \frac{1}{2 \omega_k} e^{+ik \cdot (x - y)} \Big|_{k_0 = \omega_k} ~, \notag \\[1em]
\text{Pauli-Jordan Function: } \hspace{1.3cm} &\notag\\[0.5em]
\Delta_{xy}^\phi \equiv \Delta^\phi (x-y) &= \comm\big{\phi (x)}{\phi (y)} \label{eq:PJfunction} \\
&= \int \frac{\dd[3]{\mathbf{k}}}{(2\pi)^3} \frac{1}{2 \omega_k} \bigl( e^{-ik \cdot (x - y)} - e^{ik \cdot (x - y)} \bigr) \Big|_{k_0 = \omega_k} ~, \notag \\[1em]
\text{Hadamard Function: } \hspace{1.6cm}&\notag\\[0.5em]
\Delta_{xy}^{\phi(H)} \equiv \Delta^{\phi(H)} (x-y) &=  \bra{0^\phi}\acomm\big{\phi (x)}{\phi (y)}\ket{0^\phi} \label{eq:Hadamardfunction}\\
&=  \int \frac{\dd[3]{\mathbf{k}}}{(2\pi)^3} \frac{1}{2 \omega_k} \bigl( e^{-ik \cdot (x - y)} + e^{ik \cdot (x - y)} \bigr) \Big|_{k_0 = \omega_k} \notag\\
& = 2 \bra{0^\phi} \phi (x) \phi (y) \ket{0^\phi} - \Delta_{xy}^{\phi}  ~.\notag\\[1em]
\text{Feynman Propagator: } \hspace{1.4cm}&\notag\\[0.5em]
F_{xy}^\phi \equiv F^\phi (x-y) &= \expval**{\text{T} \big\{ \phi (x) \phi (y) \big\}}{0^\phi} \label{eq:FeynmanDefinition} \\
&= \int \frac{\dd[3]{\mathbf{k}}}{(2\pi)^3} \frac{1}{2 \omega_k} \bigl( \Theta (x_0 - y_0) e^{-ik \cdot (x - y)} \notag \\
&\quad \quad+ \Theta (y_0 - x_0) e^{ik \cdot (x - y)} \bigr) \Big|_{k_0 = \omega_k} \notag \\
&= i \int \frac{\dd[4]{k}}{(2\pi)^4} \frac{e^{-ik \cdot (x - y)}}{k^2 - m^2 + i \epsilon} \,, \notag \\[1em]
\text{Retarded Propagator: } \hspace{1.4cm}&\notag\\[0.5em]
R_{xy}^\phi \equiv R^\phi (x-y) &= \Theta (x_0 - y_0) \Delta (x-y) \label{eq:RetardedDefinition} \\
&= \Theta (x_0 - y_0) \comm\big{\phi (x)}{\phi (y)} \notag \\
&= i \int \frac{\dd[4]{k}}{(2\pi)^4} \frac{e^{-ik \cdot (x - y)}}{k^2 - m^2 + k_0 i \epsilon} \,. \notag
\end{align}
}

\section{First-Order Decay Expressions}\label{app:firstorderdecay}

Eqs.~\eqref{eq:expval1}--\eqref{eq:expval12} display the full expressions of each term in Eq.~\eqref{eq:F4expvalexpanded}. 

{ \allowdisplaybreaks
\begin{align*}
    \langle E^h_{\ul{2}3}\,E^\chi_{14}\,\mathcal{E}^{\chi h h \chi}_{\ul{1}2 \ul{3}\ul{4}} \rangle =&\, 2 \,\Delta^{h}_{23} \,\frac{1}{2 \omega_p} \bigl( - e^{i p \cdot x_1} \, e^{-i p \cdot x_4} - \, e^{i p \cdot x_4} \, e^{-i p \cdot x_1} \bigr) \times 16 \,g_h^2 g_\chi^2\,\Delta^\phi_{12} \\
    & \bra{0^\phi} \Biggl( 2\,  \,\phi_1 \,\phi_2 \,\phi_3^2 \,\phi_4^2 -  \,\Delta^\phi_{12} \,\phi_3^2 \,\phi_4^2  \\
    & \hspace{9mm}  - 2\,\Delta^\phi_{13}  \,\phi_2 \,\phi_3 \,\phi_4^2 - 2\,\Delta^\phi_{23} \,\phi_1 \,\phi_3 \,\phi_4^2  + 2\,\Delta^\phi_{13}\,\Delta^\phi_{23} \,\phi_4^2  
      \\
    & \hspace{9mm} - 2\,\Delta^\phi_{14} 
    \biggl[    \,\phi_2\,\phi_3^2\,\phi_4  - \Delta^\phi_{23} \,\phi_3\,\phi_4
    \biggr]   - 2\,\Delta^\phi_{24} 
    \biggl[   \,\phi_1\,\phi_3^2\,\phi_4  - \Delta^\phi_{13}\, \,\phi_3\,\phi_4
    \biggr] \\
    & \hspace{9mm} - 2\,\Delta^\phi_{34} 
    \biggl[  2\,  \,\phi_1\,\phi_2\,\phi_3\,\phi_4 -  \Delta^\phi_{12} \,\phi_3 \,\phi_4  - \Delta^\phi_{13}  \,\phi_2\,\phi_4 - \Delta^\phi_{23}  \,\phi_1\,\phi_4
    \biggr] \\
    & \hspace{9mm} +2\,\Delta^\phi_{14}\,\Delta^\phi_{24} 
      \,\phi_3^2 +2\,\Delta^\phi_{14}\,\Delta^\phi_{34} 
    \biggl[ 2\, \,\phi_2\,\phi_3 - \,\Delta^\phi_{23}   
    \biggr] \\
    & \hspace{9mm} +2\,\Delta^\phi_{24}\,\Delta^\phi_{34} 
    \biggl[ 2\, \,\phi_1\,\phi_3  - \,\Delta^\phi_{13}   
    \biggr]  + (\Delta^\phi_{34})^2
    \biggl[ 2\, \,\phi_1\,\phi_2 - \Delta^\phi_{12}
    \biggr]
    \Biggr) \ket{0^\phi} \numberthis
    \label{eq:expval1}
\end{align*}
\begin{align*}
    \langle E^h_{\ul{2}3}\,E^\chi_{1\ul{4}}\,\mathcal{E}^{\chi h h \chi}_{\ul{1}2 \ul{3}4} \rangle =&\, 2 \,\Delta^{h}_{23} \,\frac{1}{2 \omega_p} \bigl( - e^{i p \cdot x_1} \, e^{-i p \cdot x_4} + \, e^{i p \cdot x_4} \, e^{-i p \cdot x_1} \bigr) \times 16\,g_h^2 g_\chi^2\,\Delta^\phi_{12} \\
    & \bra{0^\phi} \Biggl( 2\,\Delta^\phi_{14} 
    \biggl[   \,\phi_2\,\phi_3^2\,\phi_4 -\Delta^\phi_{23}\, \,\phi_3\,\phi_4
    \biggr] + 2\,\Delta^\phi_{24} 
    \biggl[   \,\phi_1\,\phi_3^2\,\phi_4 - \Delta^\phi_{13}\, \,\phi_3\,\phi_4 
    \biggr] \\
    & \hspace{9mm} + 2\,\Delta^\phi_{34} 
    \biggl[ 2\,  \,\phi_1\,\phi_2\,\phi_3\,\phi_4 -  \Delta^\phi_{12} \,\phi_3 \,\phi_4 - \Delta^\phi_{13}  \,\phi_2\,\phi_4 
    - \Delta^\phi_{23}  \,\phi_1\,\phi_4 
    \biggr] \\
    & \hspace{9mm} -2\,\Delta^\phi_{14}\,\Delta^\phi_{24}\, \,\phi_3^2  -2\,\Delta^\phi_{14}\,\Delta^\phi_{34} 
    \biggl[ 2\, \,\phi_2\,\phi_3 - \Delta^\phi_{23} \, 
    \biggr] \\
    & \hspace{9mm} -2\,\Delta^\phi_{24}\,\Delta^\phi_{34} 
    \biggl[  2\, \,\phi_1\,\phi_3 - \Delta^\phi_{13}\, 
    \biggr] -(\Delta^\phi_{34})^2
    \biggl[2\, \,\phi_1\,\phi_2- \Delta^\phi_{12} 
    \biggr]
    \Biggr) \ket{0^\phi} \numberthis
    \label{eq:expval2}
\end{align*}
\begin{align*}
    \langle E^h_{\ul{2}\ul{3}}\,E^\chi_{14}\,\mathcal{E}^{\chi h h \chi}_{\ul{1}2 3\ul{4}} \rangle =&\, 2 \,\Delta^{h(H)}_{23} \,\frac{1}{2 \omega_p} \bigl( - e^{i p \cdot x_1} \, e^{-i p \cdot x_4} - \, e^{i p \cdot x_4} \, e^{-i p \cdot x_1} \bigr) \times 32 \,g_h^2 g_\chi^2 \,\Delta^\phi_{12} \\
    & \bra{0^\phi} \Biggl( \Delta^\phi_{13} \phi_2 \,\phi_3 \,\phi_4^2 + \Delta^\phi_{23} \,\phi_1 \,\phi_3 \,\phi_4^2 - \Delta^\phi_{13}\,\Delta^\phi_{23}\,\phi_4^2  \\
    & \hspace{12mm}  - \Delta^\phi_{14} 
    \Delta^\phi_{23}\,\phi_3\,\phi_4 - \Delta^\phi_{24} 
    \Delta^\phi_{13}\,\phi_3\,\phi_4 \\
    & \hspace{12mm} - \Delta^\phi_{34} 
    \biggl[  \Delta^\phi_{13}\,\phi_2\,\phi_4 + \Delta^\phi_{23} \,\phi_1\,\phi_4 -\Delta^\phi_{14}\,\Delta^\phi_{23} -\Delta^\phi_{24}\,\Delta^\phi_{13} \biggr]
    \Biggr) \ket{0^\phi} \numberthis
    \label{eq:expval3}
\end{align*}
\begin{align*}
    \langle E^h_{\ul{2}\ul{3}}\,E^\chi_{1\ul{4}}\,\mathcal{E}^{\chi h h \chi}_{\ul{1}2 3 4} \rangle = &\, 2 \,\Delta^{h(H)}_{23} \,\frac{1}{2 \omega_p} \bigl( - e^{i p \cdot x_1} \, e^{-i p \cdot x_4} + \, e^{i p \cdot x_4} \, e^{-i p \cdot x_1} \bigr) \times 32\,g_h^2 g_\chi^2 \,\Delta^\phi_{12} \\
    &\bra{0^\phi} \Biggl( \Delta^\phi_{14} 
    \Delta^\phi_{23}\,\phi_3\,\phi_4 + \Delta^\phi_{24} 
    \Delta^\phi_{13}\,\phi_3\,\phi_4 \\
    & \hspace{12mm} + \Delta^\phi_{34} 
    \biggl[  \Delta^\phi_{13}\,\phi_2\,\phi_4 + \Delta^\phi_{23} \,\phi_1\,\phi_4 -\Delta^\phi_{14}\,\Delta^\phi_{23} -\Delta^\phi_{24}\,\Delta^\phi_{13} \biggr]
    \Biggr) \ket{0^\phi} \numberthis
    \label{eq:expval4}
\end{align*}
\begin{align*}
    \langle E^h_{\ul{2}4}\,E^\chi_{13}\,\mathcal{E}^{\chi h \chi h}_{\ul{1}2 \ul{3}\ul{4}} \rangle = &\, 2 \,\Delta^{h}_{24} \,\frac{1}{2 \omega_p} \bigl( - e^{i p \cdot x_1} \, e^{-i p \cdot x_3} - \, e^{i p \cdot x_3} \, e^{-i p \cdot x_1} \bigr) \times 16\,g_h^2 g_\chi^2 \,\Delta^\phi_{12} \\
    & \bra{0^\phi} \Biggl( 2\,  \,\phi_1 \,\phi_2 \,\phi_3^2 \,\phi_4^2 -  \,\Delta^\phi_{12} \,\phi_3^2 \,\phi_4^2  \\
    & \hspace{5mm}  - 2\,\Delta^\phi_{13}  \,\phi_2 \,\phi_3 \,\phi_4^2 - 2\,\Delta^\phi_{23} \,\phi_1 \,\phi_3 \,\phi_4^2  + 2\,\Delta^\phi_{13}\,\Delta^\phi_{23} \,\phi_4^2  
      \\
    & \hspace{5mm} - 2\,\Delta^\phi_{14} 
    \biggl[    \,\phi_2\,\phi_3^2\,\phi_4  - \Delta^\phi_{23} \,\phi_3\,\phi_4
    \biggr]   - 2\,\Delta^\phi_{24} 
    \biggl[   \,\phi_1\,\phi_3^2\,\phi_4  - \Delta^\phi_{13}\, \,\phi_3\,\phi_4
    \biggr] \\
    & \hspace{5mm} - 2\,\Delta^\phi_{34} 
    \biggl[  2\,  \,\phi_1\,\phi_2\,\phi_3\,\phi_4 -  \Delta^\phi_{12} \,\phi_3 \,\phi_4  - \Delta^\phi_{13}  \,\phi_2\,\phi_4 - \Delta^\phi_{23}  \,\phi_1\,\phi_4
    \biggr] \\
    & \hspace{5mm} +2\,\Delta^\phi_{14}\,\Delta^\phi_{24} 
      \,\phi_3^2 +2\,\Delta^\phi_{14}\,\Delta^\phi_{34} 
    \biggl[ 2\, \,\phi_2\,\phi_3 - \,\Delta^\phi_{23}   
    \biggr] \\
    & \hspace{5mm} +2\,\Delta^\phi_{24}\,\Delta^\phi_{34} 
    \biggl[ 2\, \,\phi_1\,\phi_3  - \,\Delta^\phi_{13}   
    \biggr]  + (\Delta^\phi_{34})^2
    \biggl[ 2\, \,\phi_1\,\phi_2 - \Delta^\phi_{12}
    \biggr]
    \Biggr) \ket{0^\phi} \numberthis
    \label{eq:expval5}
\end{align*}
\begin{align*}
    \langle E^h_{\ul{2}4}\,E^\chi_{1\ul{3}}\,\mathcal{E}^{\chi h \chi h}_{\ul{1}2 3\ul{4}} \rangle = &\, 2 \,\Delta^{h}_{24} \,\frac{1}{2 \omega_p} \bigl( - e^{i p \cdot x_1} \, e^{-i p \cdot x_3} + \, e^{i p \cdot x_3} \, e^{-i p \cdot x_1} \bigr) \times  32\,g_h^2 g_\chi^2 \,\Delta^\phi_{12} \\
    &\bra{0^\phi} \Biggl( \Delta^\phi_{13} \phi_2 \,\phi_3 \,\phi_4^2 + \Delta^\phi_{23} \,\phi_1 \,\phi_3 \,\phi_4^2 - \Delta^\phi_{13}\,\Delta^\phi_{23}\,\phi_4^2  \\
    & \hspace{12mm}  - \Delta^\phi_{14} 
    \Delta^\phi_{23}\,\phi_3\,\phi_4 - \Delta^\phi_{24} 
    \Delta^\phi_{13}\,\phi_3\,\phi_4 \\
    & \hspace{12mm} - \Delta^\phi_{34} 
    \biggl[  \Delta^\phi_{13}\,\phi_2\,\phi_4 + \Delta^\phi_{23} \,\phi_1\,\phi_4 -\Delta^\phi_{14}\,\Delta^\phi_{23} -\Delta^\phi_{24}\,\Delta^\phi_{13} \biggr]
    \Biggr) \ket{0^\phi} \numberthis
    \label{eq:expval6}
\end{align*}
\begin{align*}
    \langle E^h_{\ul{2}\ul{4}}\,E^\chi_{13}\,\mathcal{E}^{\chi h \chi h}_{\ul{1}2 \ul{3}4} \rangle = &\, 2 \,\Delta^{h(H)}_{24} \,\frac{1}{2 \omega_p} \bigl( - e^{i p \cdot x_1} \, e^{-i p \cdot x_3} - \, e^{i p \cdot x_3} \, e^{-i p \cdot x_1} \bigr) \times 16\,g_h^2 g_\chi^2\,\Delta^\phi_{12} \\
    & \bra{0^\phi} \Biggl( 2\,\Delta^\phi_{14} 
    \biggl[   \,\phi_2\,\phi_3^2\,\phi_4 -\Delta^\phi_{23}\, \,\phi_3\,\phi_4
    \biggr] + 2\,\Delta^\phi_{24} 
    \biggl[   \,\phi_1\,\phi_3^2\,\phi_4 - \Delta^\phi_{13}\, \,\phi_3\,\phi_4 
    \biggr] \\
    & \hspace{9mm} + 2\,\Delta^\phi_{34} 
    \biggl[ 2\,  \,\phi_1\,\phi_2\,\phi_3\,\phi_4 -  \Delta^\phi_{12} \,\phi_3 \,\phi_4 - \Delta^\phi_{13}  \,\phi_2\,\phi_4 
    - \Delta^\phi_{23}  \,\phi_1\,\phi_4 
    \biggr] \\
    & \hspace{9mm} -2\,\Delta^\phi_{14}\,\Delta^\phi_{24}\, \,\phi_3^2  -2\,\Delta^\phi_{14}\,\Delta^\phi_{34} 
    \biggl[ 2\, \,\phi_2\,\phi_3 - \Delta^\phi_{23} \, 
    \biggr] \\
    & \hspace{9mm} -2\,\Delta^\phi_{24}\,\Delta^\phi_{34} 
    \biggl[  2\, \,\phi_1\,\phi_3 - \Delta^\phi_{13}\, 
    \biggr] -(\Delta^\phi_{34})^2
    \biggl[2\, \,\phi_1\,\phi_2- \Delta^\phi_{12} 
    \biggr]
    \Biggr) \ket{0^\phi} \numberthis
    \label{eq:expval7}
\end{align*}
\begin{align*}
    \langle E^h_{\ul{2}\ul{4}}\,E^\chi_{1\ul{3}}\,\mathcal{E}^{\chi h h \chi}_{\ul{1}2 3 4} \rangle = &\, 2 \,\Delta^{h(H)}_{24} \,\frac{1}{2 \omega_p} \bigl( - e^{i p \cdot x_1} \, e^{-i p \cdot x_3} + \, e^{i p \cdot x_3} \, e^{-i p \cdot x_1} \bigr) \times 32\,g_h^2 g_\chi^2 \,\Delta^\phi_{12} \\
    &\bra{0^\phi} \Biggl( \Delta^\phi_{14} 
    \Delta^\phi_{23}\,\phi_3\,\phi_4 + \Delta^\phi_{24} 
    \Delta^\phi_{13}\,\phi_3\,\phi_4 \\
    & \hspace{12mm} + \Delta^\phi_{34} 
    \biggl[  \Delta^\phi_{13}\,\phi_2\,\phi_4 + \Delta^\phi_{23} \,\phi_1\,\phi_4 -\Delta^\phi_{14}\,\Delta^\phi_{23} -\Delta^\phi_{24}\,\Delta^\phi_{13} \biggr]
    \Biggr) \ket{0^\phi} \numberthis
    \label{eq:expval8}
\end{align*}
\begin{align*}
    \langle E^h_{\ul{3}4}\,E^\chi_{12}\,\mathcal{E}^{\chi \chi h h }_{\ul{1}\ul{2} 3\ul{4}} \rangle =&\, 2 \,\Delta^{h}_{34} \,\frac{1}{2 \omega_p} \bigl( - e^{i p \cdot x_1} \, e^{-i p \cdot x_2} - \, e^{i p \cdot x_2} \, e^{-i p \cdot x_1} \bigr) \times 16\,g_h^2 g_\chi^2 \\
    & \bra{0^\phi} \Biggl( 2\,\Delta^\phi_{13} \Bigl(  \,\phi_1 \,\phi_2^2 \,\phi_3 \,\phi_4^2 - \,\Delta^\phi_{12} \,\phi_2 \,\phi_3 \,\phi_4^2 \Bigr)  \\ 
    & \hspace{11mm} + 2\,\Delta^\phi_{23} \Bigl(  \phi_1^2 \,\phi_2 \,\phi_3 \,\phi_4^2 -   \,\Delta^\phi_{12} \,\phi_1 \,\phi_3 \,\phi_4^2 \Bigr)  \\
    & \hspace{11mm} - 2\,\Delta^\phi_{13}\,\Delta^\phi_{23} \Bigl(  2 \,\phi_1\,\phi_2\,\phi_4^2  -   \,\Delta^\phi_{12}\,\phi_4^2\Bigr)  \\
    & \hspace{11mm} -  (\Delta^\phi_{13})^2 \,\phi_2^2 \,\phi_4^2 - (\Delta^\phi_{23})^2 \phi_1^2 \,\phi_4^2  \\
    & \hspace{11mm} + 2\,\Delta^\phi_{14} 
    \biggl[ - \Delta^\phi_{13}\,\phi_2^2\,\phi_3\,\phi_4 - \Delta^\phi_{23}\bigl( 2\,\phi_1\,\phi_2\,\phi_3\,\phi_4-  \,\Delta^\phi_{12}\,\phi_3\,\phi_4\bigr) \\
    & \hspace{24mm} + 2 \,\Delta^\phi_{13}\,\Delta^\phi_{23}\,\phi_2\,\phi_4 + (\Delta^\phi_{23})^2\,\phi_1 \,\phi_4 
    \biggr]  \\
    & \hspace{11mm} +2\,\Delta^\phi_{24} 
    \biggl[ -  \Delta^\phi_{23}\,\phi_1^2\,\phi_3\,\phi_4  - \Delta^\phi_{13}\bigl( 2\,\phi_1\,\phi_2\,\phi_3\,\phi_4-  \,\Delta^\phi_{12}\,\phi_3\,\phi_4\bigr)\\
    & \hspace{27mm} + 2 \,\Delta^\phi_{13}\,\Delta^\phi_{23}\,\phi_1\,\phi_4 +  
    (\Delta^\phi_{13})^2\,\phi_2 \,\phi_4
    \biggr] \\
    & \hspace{11mm} +2 \,\Delta^\phi_{34} 
    \biggl[  - \Delta^\phi_{13}\bigl(  \phi_1\,\phi_2^2\,\phi_4-  \,\Delta^\phi_{12}\,\phi_2\,\phi_4\bigr) \\
    & \hspace{27mm} - \Delta^\phi_{23}\bigl(  \phi_1^2\,\phi_2\,\phi_4 -  \,\Delta^\phi_{12}\,\phi_1\,\phi_4\bigr)
    \biggr] \\
    & \hspace{11mm} +4\,\Delta^\phi_{14}\,\Delta^\phi_{24} 
    \biggl[ \Delta^\phi_{13}\,\phi_2\,\phi_3 + \Delta^\phi_{23} \,\phi_1 \,\phi_3 - \Delta^\phi_{13} \, \Delta^\phi_{23}
    \biggr] \\
    & \hspace{11mm} +2\,\Delta^\phi_{14}\,\Delta^\phi_{34} 
    \biggl[  \Delta^\phi_{13}\,\phi_2^2 + \Delta^\phi_{23} \bigl(  2 \,\phi_1 \,\phi_2 -  \,\Delta^\phi_{12} \bigr)
    \biggr] \\
    & \hspace{11mm} + 2 \,\Delta^\phi_{24}\,\Delta^\phi_{34} 
    \biggl[  \Delta^\phi_{23}\,\phi_1^2 + \Delta^\phi_{13} \bigl(  2 \,\phi_1 \,\phi_2 -  \,\Delta^\phi_{12} \bigr)
    \biggr] \\
    & \hspace{11mm} +  (\Delta^\phi_{14})^2
    \biggl[ 2\,\Delta^\phi_{23}\,\phi_2\,\phi_3 - (\Delta^\phi_{23})^2 
    \biggr]
    \\
    & \hspace{11mm} +  (\Delta^\phi_{24})^2
    \biggl[ 2\,\Delta^\phi_{13}\,\phi_1\,\phi_3 - (\Delta^\phi_{13})^2
    \biggr]
    \Biggr) \ket{0^\phi} \numberthis
    \label{eq:expval9}
\end{align*}
\begin{align*}
    \langle E^h_{\ul{3}4}\,E^\chi_{1\ul{2}}\,\mathcal{E}^{\chi \chi h h }_{\ul{1}2 3\ul{4}} \rangle =&\, 2 \,\Delta^{h}_{34} \,\frac{1}{2 \omega_p} \bigl( - e^{i p \cdot x_1} \, e^{-i p \cdot x_2} + \, e^{i p \cdot x_2} \, e^{-i p \cdot x_1} \bigr) \times 32\,g_h^2 g_\chi^2 \,\Delta^\phi_{12} \\
    &\bra{0^\phi} \Biggl( \Delta^\phi_{13} \phi_2 \,\phi_3 \,\phi_4^2 + \Delta^\phi_{23} \,\phi_1 \,\phi_3 \,\phi_4^2 - \Delta^\phi_{13}\,\Delta^\phi_{23}\,\phi_4^2  \\
    & \hspace{12mm}  - \Delta^\phi_{14} 
    \Delta^\phi_{23}\,\phi_3\,\phi_4 - \Delta^\phi_{24} 
    \Delta^\phi_{13}\,\phi_3\,\phi_4 \\
    & \hspace{12mm} - \Delta^\phi_{34} 
    \biggl[  \Delta^\phi_{13}\,\phi_2\,\phi_4 + \Delta^\phi_{23} \,\phi_1\,\phi_4 -\Delta^\phi_{14}\,\Delta^\phi_{23} -\Delta^\phi_{24}\,\Delta^\phi_{13} \biggr]
    \Biggr) \ket{0^\phi} \numberthis
    \label{eq:expval10}
\end{align*}
\begin{align*}
    \langle E^h_{\ul{3}\ul{4}}\,E^\chi_{12}\,\mathcal{E}^{\chi \chi h h }_{\ul{1}\ul{2} 3 4} \rangle =&\, 2 \,\Delta^{h(H)}_{34} \,\frac{1}{2 \omega_p} \bigl( - e^{i p \cdot x_1} \, e^{-i p \cdot x_2} - \, e^{i p \cdot x_2} \, e^{-i p \cdot x_1} \bigr) \times 16\,g_h^2 g_\chi^2\, \\
    & \bra{0^\phi} \Biggl( 2\,\Delta^\phi_{14} 
    \biggl[  \Delta^\phi_{13}\,\phi_2^2\,\phi_3\,\phi_4  + \Delta^\phi_{23}\bigl(2\,\phi_1\,\phi_2\,\phi_3\,\phi_4-  \,\Delta^\phi_{12}\,\phi_3\,\phi_4\bigr) \\
    & \hspace{23mm} -2 \,\Delta^\phi_{13}\,\Delta^\phi_{23}\,\phi_2\,\phi_4 - (\Delta^\phi_{23})^2\,\phi_1 \,\phi_4 
    \biggr]  \\
    & \hspace{12mm} 2\,\Delta^\phi_{24} 
    \biggl[  \Delta^\phi_{23}\,\phi_1^2\,\phi_3\,\phi_4  + \Delta^\phi_{13}\bigl(2\,\phi_1\,\phi_2\,\phi_3\,\phi_4-  \,\Delta^\phi_{12}\,\phi_3\,\phi_4\bigr)\\
    & \hspace{23mm} -2 \,\Delta^\phi_{13}\,\Delta^\phi_{23}\,\phi_1\,\phi_4 - (\Delta^\phi_{13})^2\,\phi_2 \,\phi_4 \Bigr)
    \biggr] \\
    & \hspace{12mm} 2\,\Delta^\phi_{34} 
    \biggl[ \Delta^\phi_{13}\bigl( \phi_1\,\phi_2^2\,\phi_4-  \,\Delta^\phi_{12}\,\phi_2\,\phi_4\bigr) \\
    & \hspace{23mm} + \Delta^\phi_{23}\bigl( \phi_1^2\,\phi_2\,\phi_4-  \,\Delta^\phi_{12}\,\phi_1\,\phi_4\bigr) 
    \biggr] \\
    & \hspace{12mm} -4\,\Delta^\phi_{14}\,\Delta^\phi_{24} 
    \biggl[ \Delta^\phi_{13}\,\phi_2\,\phi_3 + \Delta^\phi_{23} \,\phi_1 \,\phi_3 - \,\Delta^\phi_{13} \, \Delta^\phi_{23} 
    \biggr] \\
    & \hspace{12mm} -2\,\Delta^\phi_{14}\,\Delta^\phi_{34} 
    \biggl[  \Delta^\phi_{13}\,\phi_2^2 + \Delta^\phi_{23} \bigl( 2 \,\phi_1 \,\phi_2 -  \,\Delta^\phi_{12} \bigr)
    \biggr] \\
    & \hspace{12mm} -2\,\Delta^\phi_{24}\,\Delta^\phi_{34} 
    \biggl[ \Delta^\phi_{23}\,\phi_1^2 + \Delta^\phi_{13} \bigl( 2 \,\phi_1 \,\phi_2 -  \,\Delta^\phi_{12} \bigr) 
    \biggr] \\
    & \hspace{12mm} - (\Delta^\phi_{14})^2
    \biggl[ 2\,\Delta^\phi_{23}\,\phi_2\,\phi_3-(\Delta^\phi_{23})^2 
    \biggr]
    \\
    & \hspace{12mm} - (\Delta^\phi_{24})^2
    \biggl[ 2\,\Delta^\phi_{13}\,\phi_1\,\phi_3-(\Delta^\phi_{13})^2 
    \biggr]
    \Biggr) \ket{0^\phi} \numberthis
    \label{eq:expval11}
\end{align*}
\begin{align*}
    \langle E^h_{\ul{3}\ul{4}}\,E^\chi_{1\ul{2}}\,\mathcal{E}^{\chi \chi h h }_{\ul{1}2 3 4} \rangle =&\, 2 \,\Delta^{h(H)}_{34} \,\frac{1}{2 \omega_p} \bigl( - e^{i p \cdot x_1} \, e^{-i p \cdot x_2} + \, e^{i p \cdot x_2} \, e^{-i p \cdot x_1} \bigr) \times 32\,g_h^2 g_\chi^2 \,\Delta^\phi_{12} \\
    &\bra{0^\phi} \Biggl( \Delta^\phi_{14} 
    \Delta^\phi_{23}\,\phi_3\,\phi_4 + \Delta^\phi_{24} 
    \Delta^\phi_{13}\,\phi_3\,\phi_4 \\
    & \hspace{12mm} + \Delta^\phi_{34} 
    \biggl[  \Delta^\phi_{13}\,\phi_2\,\phi_4 + \Delta^\phi_{23} \,\phi_1\,\phi_4 -\Delta^\phi_{14}\,\Delta^\phi_{23} -\Delta^\phi_{24}\,\Delta^\phi_{13} \biggr]
    \Biggr) \ket{0^\phi} \numberthis
    \label{eq:expval12}
\end{align*}
}

\section{Symmetrising the Time-Ordering in the Annihilation Probability}\label{app:timesymmetry}

Here, we show analytically how the time-ordered integral in Eq.~\eqref{eq:expval2to2} can be symmetrised and written as an integral over all times, as is shown in Fig.~\ref{fig:timesymmetry}. We only show this for the diagrams containing Feynman loops, but the same procedure can be applied to each line in Fig.~\ref{fig:timesymmetry}.

The terms containing Feynman loops in the annihilation probability are
\begin{align}\label{eq:feynmanloopswithTheta}
    \mathbb{P} \supset -\frac{4 \,g_\psi^2 g_\phi^2}{   (2\omega_{p_1})(2\omega_{p_2})} \int & \D^4 {x_{1}} \,\D^4 {x_{2}} \,\D^4 {x_{3}} \,\D^4 {x_{4}}  \nonumber \\
    \Biggl( &e^{ip_1\cdot x_1}e^{ip_2\cdot x_1}e^{-ip_1\cdot x_2}e^{-ip_2\cdot x_2} 
    \left( F_{34}^{\phi}\right)^2 R_{13}^\chi R_{24}^\chi \,\Theta_{12}\Theta_{23}\Theta_{34}\nonumber\\
    +\,&e^{ip_1\cdot x_1}e^{ip_2\cdot x_1}e^{-ip_1\cdot x_2}e^{-ip_2\cdot x_2} 
    \left( F_{34}^{\phi}\right)^2 R_{14}^\chi R_{23}^\chi \,\Theta_{12}\Theta_{23}\Theta_{34}\nonumber\\
    +\,&e^{ip_1\cdot x_1}e^{ip_2\cdot x_1}e^{-ip_1\cdot x_3}e^{-ip_2\cdot x_3} 
    \left( F_{24}^{\phi}\right)^2 R_{12}^\chi R_{34}^\chi \,\Theta_{12}\Theta_{23}\Theta_{34}\nonumber\\
    +\,&e^{-ip_1\cdot x_1}e^{-ip_2\cdot x_1}e^{ip_1\cdot x_2}e^{ip_2\cdot x_2} 
    \left( F_{34}^{\phi}\right)^2 R_{13}^\chi R_{24}^\chi \,\Theta_{12}\Theta_{23}\Theta_{34}\nonumber\\
    +\,&e^{-ip_1\cdot x_1}e^{-ip_2\cdot x_1}e^{ip_1\cdot x_2}e^{ip_2\cdot x_2} 
    \left( F_{34}^{\phi}\right)^2 R_{14}^\chi R_{23}^\chi \,\Theta_{12}\Theta_{23}\Theta_{34}\nonumber\\
    +\,&e^{-ip_1\cdot x_1}e^{-ip_2\cdot x_1}e^{ip_1\cdot x_3}e^{ip_2\cdot x_3} 
    \left( F_{24}^{\phi}\right)^2 R_{12}^\chi R_{34}^\chi \,\Theta_{12}\Theta_{23}\Theta_{34}
     \Biggr)
    ~,
\end{align}
where the $\Theta$-function has been rewritten as $\Theta_{1234} = \Theta_{12}\Theta_{23}\Theta_{34}$. We are free to re-label the integration variables in each term as
\begin{align}
    \mathbb{P} \supset -\frac{4 \,g_\psi^2 g_\phi^2}{   (2\omega_{p_1})(2\omega_{p_2})} \int & \D^4 {x_{1}} \,\D^4 {x_{2}} \,\D^4 {x_{3}} \,\D^4 {x_{4}}  \nonumber \\
    \Biggl( &e^{ip_1\cdot x_1}e^{ip_2\cdot x_1}e^{-ip_1\cdot x_2}e^{-ip_2\cdot x_2} 
    \left( F_{34}^{\phi}\right)^2 R_{13}^\chi R_{24}^\chi \,\Theta_{12}\Theta_{23}\Theta_{34}\nonumber\\
    +\,&e^{ip_1\cdot x_1}e^{ip_2\cdot x_1}e^{-ip_1\cdot x_2}e^{-ip_2\cdot x_2} 
    \left( F_{34}^{\phi}\right)^2 R_{13}^\chi R_{24}^\chi \,\Theta_{12}\Theta_{24}\Theta_{43}\nonumber\\
    +\,&e^{ip_1\cdot x_1}e^{ip_2\cdot x_1}e^{-ip_1\cdot x_2}e^{-ip_2\cdot x_2} 
    \left( F_{34}^{\phi}\right)^2 R_{13}^\chi R_{24}^\chi \,\Theta_{13}\Theta_{32}\Theta_{24}\nonumber\\
    +\,&e^{-ip_1\cdot x_1}e^{-ip_2\cdot x_1}e^{ip_1\cdot x_2}e^{ip_2\cdot x_2} 
    \left( F_{34}^{\phi}\right)^2 R_{14}^\chi R_{23}^\chi \,\Theta_{12}\Theta_{24}\Theta_{43}\nonumber\\
    +\,&e^{-ip_1\cdot x_1}e^{-ip_2\cdot x_1}e^{ip_1\cdot x_2}e^{ip_2\cdot x_2} 
    \left( F_{34}^{\phi}\right)^2 R_{14}^\chi R_{23}^\chi \,\Theta_{12}\Theta_{23}\Theta_{34}\nonumber\\
    +\,&e^{-ip_1\cdot x_1}e^{-ip_2\cdot x_1}e^{ip_1\cdot x_2}e^{ip_2\cdot x_2} 
    \left( F_{34}^{\phi}\right)^2 R_{14}^\chi R_{23}^\chi \,\Theta_{14}\Theta_{42}\Theta_{23}
     \Biggr)
    ~,
\end{align}
where we have used $F_{xy}^\phi = F_{yx}^\phi$. Using the relations 
\begin{align}
    \Theta_{xy}\Theta_{yz} &= \Theta_{xy}\Theta_{yz}\Theta_{xz} \\
    \Theta_{xy} + \Theta_{yx} &= 1 \\
    \Theta_{xy} R_{xy}^\chi &= R_{xy}^\chi \,,
\end{align}
we obtain
\begin{align}
    \mathbb{P} \supset -\frac{4 \,g_\psi^2 g_\phi^2}{   (2\omega_{p_1})(2\omega_{p_2})} \int & \D^4 {x_{1}} \,\D^4 {x_{2}} \,\D^4 {x_{3}} \,\D^4 {x_{4}}  \nonumber \\
    \Biggl( &e^{ip_1\cdot x_1}e^{ip_2\cdot x_1}e^{-ip_1\cdot x_2}e^{-ip_2\cdot x_2} 
    \left( F_{34}^{\phi}\right)^2 R_{13}^\chi R_{24}^\chi \,\Theta_{12}\nonumber\\
    +\,&e^{-ip_1\cdot x_1}e^{-ip_2\cdot x_1}e^{ip_1\cdot x_2}e^{ip_2\cdot x_2} 
    \left( F_{34}^{\phi}\right)^2 R_{14}^\chi R_{23}^\chi \,\Theta_{12}
     \Biggr)
    ~.
\end{align}
Re-labelling $x_1 \leftrightarrow x_2$ in the second term yields
\begin{align}
    \mathbb{P} \supset -\frac{4 \,g_\psi^2 g_\phi^2}{   (2\omega_{p_1})(2\omega_{p_2})} \int & \D^4 {x_{1}} \,\D^4 {x_{2}} \,\D^4 {x_{3}} \,\D^4 {x_{4}}  \nonumber \\
    &e^{ip_1\cdot x_1}e^{ip_2\cdot x_1}e^{-ip_1\cdot x_2}e^{-ip_2\cdot x_2} 
    \left( F_{34}^{\phi}\right)^2 R_{13}^\chi R_{24}^\chi \left( \Theta_{12} + \Theta_{21} \right) \nonumber\\
    \supset -\frac{4 \,g_\psi^2 g_\phi^2}{   (2\omega_{p_1})(2\omega_{p_2})} \int & \D^4 {x_{1}} \,\D^4 {x_{2}} \,\D^4 {x_{3}} \,\D^4 {x_{4}}  e^{ip_1\cdot x_1}e^{ip_2\cdot x_1}e^{-ip_1\cdot x_2}e^{-ip_2\cdot x_2} 
    \left( F_{34}^{\phi}\right)^2 R_{13}^\chi R_{24}^\chi
    ~.
\end{align}
This result is equal to the first term in Eq.~\eqref{eq:feynmanloopswithTheta}, but integrated over all times since there is no time-ordering $\Theta$-function.

\bibliography{bib}

\end{document}